\documentclass[12pt]{article}
\usepackage{a4wide}
\usepackage{latexsym,amssymb,amsfonts,amsbsy,amsmath}
\usepackage{aas_macros}
\usepackage{palatino}
\usepackage[section]{placeins}
\usepackage{color}
\usepackage{graphicx}
\usepackage{cite}
\usepackage[utf8]{inputenc}
\def\vec#1{\boldsymbol{#1}}
\def\vi#1#2{\vec{#1}_#2}
\def\vis#1#2{\vec{#1}^2_#2}
\def\vs#1{\vec{#1}^2}
\def\ssspin#1#2{\boldsymbol{\sigma}_{#1}.\boldsymbol{\sigma}_{#2}}
\def\llcol#1#2{\tilde{\lambda}_{#1}.\tilde{\lambda}_{#2}}

\DeclareMathOperator{\diag}{diag} 
\DeclareMathOperator{\Tr}{Tr} 
\DeclareMathOperator{\sign}{sgn}
\input ArtNouv.fd

  \def\MM{\mathtt{M}}
 \def\TT{\mathtt{T}}
\begin{document}
%
\title{\Large{\bf HALL  POST INEQUALITIES}:\\
review and application to  molecules and tetraquarks}
\date{\small  \today}
\author{Jean-Marc~Richard\footnote{Corresponding author: \texttt{j-m.richard@ipnl.in2p3.fr}}\\[-2pt]
{\small\sl  Universit\'e de Lyon, Institut de Physique des 2  Infinis,
IN2P3-CNRS--UCBL}\\[-2pt]
{\small\sl 4 rue Enrico Fermi, 69622  Villeurbanne, France}\\[5pt]
{Alfredo~Valcarce\footnote{email: \texttt{valcarce@usal.es}}}\\[-2pt]
{\small\sl Departamento de F\'\i sica Fundamental}\\[-2pt]
{\small\sl Universidad de Salamanca, 37008 Salamanca, Spain}\\[5pt]
{Javier~Vijande\thanks{email: \texttt{javier.vijande@uv.es}}}\\[-2pt]
{\small \sl Departamento de F\'isica At\'omica, Molecular y Nuclear, Universidad de Valencia (UV)}\\[-2pt]
{\small \sl and IFIC (UV-CSIC), 46100 Valencia, Spain}\\[-3pt]
{\small \sl and}\\[-3pt]
{\small \sl IRIMED Joint Research Unit (IIS La Fe - UV), 46100 Valencia, Spain}}
%
%
 %
 \begin{titlepage}
 \maketitle
\begin{abstract}
 \noindent
  A review is presented of the Hall-Post inequalities that give lower-bounds to the ground-state energy of quantum systems in terms of energies of smaller systems. New applications are given  for systems experiencing both a static source and inner interactions, as well as for hydrogen-like molecules and for tetraquarks in some quark models. In the latter case, the Hall-Post inequalities constrain the possibility of deeply-bound exotic mesons below the threshold for dissociation into two quark-antiquark mesons.  We also emphasize the usefulness of the Hall-Post bounds in terms of 3-body energies when some 2-body subsystems are ill defined or do not support any bound state. 
\end{abstract}  
\vfil
 \end{titlepage}
\addtocounter{page}{1}
 \begin{flushright}
  \emph{Si parva licet componere magnis}\footnote{If we may compare small things with large ones (Vergilius)}
 \end{flushright}
\vfill
\newpage
 \section{Introduction}
 Hall-Post (HP)  inequalities link $N$-body energies to $N'$-body energies with $N'<N$. More precisely, the ground-state of a $N$-body system is bounded below by a sum of ground-state energies of smaller $N'$-body systems. This supplements very usefully the upper bounds provided by variational methods, and in some cases, constraints dramatically the possibility of deep binding. 
 
 There are several lower bounds to bound-state energies,  see, e.g., \cite{thirring1981quantum}. The best known is the Temple-Kato one, which for the ground state reads, 
 \begin{equation}
 \langle H\rangle_\phi - \frac{(\Delta H)_\phi}{\tilde E_1 -\langle H\rangle_\phi}\le E_0\le \langle H\rangle_\phi~,
 \end{equation}
where $\tilde E_1$ is a lower bound on the energy $E_1$ of the first excited state, provided the denominator remains positive, and $\langle H\rangle_\phi$ is a variational estimate using the normalized trial function $\phi$, which also provides the variance  $(\Delta H)_\phi=\langle H^2\rangle_\phi-\langle H\rangle^2_\phi$. We follow here the derivation by Galindo  and Pascual~\cite{Galindo:214256}. The obvious operator inequality
\begin{equation}
 (H-E_0)(H-E_1)\ge0~,
\end{equation}
translates into 
\begin{equation}
 (\Delta H)_\phi + \left(\langle H\rangle_\phi-E_0\right)\left(\langle H\rangle_\phi-E_1\right)\ge0~,
\end{equation}
from which the result follows. The Temple inequality was used for instance by Tang et al.\ \cite{1964PhRv..134..743T}, who claimed to have only 3\% difference between their variational upper bound and the associated  lower bounds for a 3-body system experiencing a short-range interaction with a strongly repulsive core.

 The Hall-Post inequalities are based on a different strategy, namely a suitable decomposition of the Hamiltonian under consideration.
 Let us give a first example of the usefulness of splitting a Hamiltonian into pieces,  considering the simple one-body operator
\begin{equation}
 H=2\, \vec{p}^2 + r^2-1/r~,
\end{equation}
with a ground state at $E\simeq 3.252$. For a simple upper bound, one can treat the Coulomb term as a perturbation: $E=E_0+E_1+\cdots$ with $E_0=3\,\sqrt2$ and $E_1=-2^{3/4}/\sqrt{\pi}$, $E_0+E_1\simeq 3.294$. If one starts from the pure Coulomb case, one gets $E=E'_0+E'_1+\cdots$ with $E'_0=-1/8$ and $E'_1=48$, thus $E'_0+E'_1=47.875$, very far, but still above the exact result.  If one wishes a lower bound,  the naive sum $H=H_1+H_2$, with $H_1=\vec{p}^2 + r^2$ of ground state $\epsilon_1$ and $H_2=\vec{p}^2 -1/r$ of ground state $\epsilon_2$,  gives a crude lower bound $E\ge \epsilon_1+\epsilon_2=2.75$, significantly below the exact result. Here, and often along this article, we use the simple result that the minimum of a sum is larger than the sum of minima. 
One can actually optimize the decomposition and write
\begin{equation}
 H=\left[(1+x)\, \vec{p}^2+r^2\right]+\left[(1-x)\,\vec{p}^2-1/r\right]~,
\end{equation}
corresponding to a lower bound
\begin{equation}
 \eta(x)=3\,(1+x)^{1/2}-(1-x)^{-1}/4~,
\end{equation}
whose maximum 
\begin{equation}
 \max_x \eta(x)\simeq 3.179~,
\end{equation}
is close to the exact value. 

This paper is devoted to decompositions of the type $H=H_1+H_2+\cdots$ where $H$ is a $N$-body Hamiltonian and the $H_i$ are $N'$-body Hamiltonians with $N'<N$, which provide lower bounds on $N$-body ground-state energies in terms of the energies of simpler systems. 
The trick was first proposed by Hall and Post to study light nuclei \cite{0370-1298-69-12-409,0370-1328-90-2-309} and developed in several papers which will be cited along the present article. The method has  been reinvented, sometimes in a degraded form, when studying the stability of matter or the relation between baryon and meson masses in the quark model \cite{1969JMP....10..806L,1982PhRvD..25.2370A,2002PhR...362..193N}. 
The aim of this article is twofold: to review the  Hall-Post inequalities, and to present some recent applications to tetraquarks in simple quark models, with the need to include coupled-channels in the formalism. Most efforts have been devoted to the case of self-interacting bosons, or, equivalently, systems of few fermions whose antisymmetry can be endorsed by the spin, isospin or color degrees of freedom, and  thus having a symmetric orbital wave function.
We shall see that the case of unequal masses requires some subtle developments, which can be applied to the case of bosons with both a pairwise interaction and the potential of a fixed source. 
The case of fermions is discussed in two paragraphs at the end of Secs.~\ref{se:naive} and \ref{se:improved}.

 In the  original formulation, the HP inequalities simply rely on the variational principle: if $H=H_1+H_2+\cdots$, then, using the ground state $\Psi$ of $H$ as a trial function immediately leads to $\min H\ge \min H_1+\min H_2+ \cdots$. For fermions, or for improving the simple Hall-Post bounds, it might be necessary to  analyze the  structure of $\Psi$, namely its content in terms of the representations of the permutation group of $(N-1)$-body clusters \cite{2001PhRvB..63g3102J} and its expansion into the tower of eigenstates of subsystems \cite{2001PhRvA..63f2107V}.  This sets the limits of the review: inequalities expressed in terms of energies of smaller systems. We shall not elaborate much on developments that require the knowledge of the wave functions of the subsystems and the solution of integro-differential equations. 
 
 Several applications will be given along this review. Originally, the HP inequalities were devised in the framework of nuclear physics, but they were applied to systems  bound by gravity, few-charge systems in atomic physics, and quark systems, on which we shall come back, with due references. Solvable models of the  Calogero-Sutherland type were considered in \cite{2001JPhA...34L.447K}.

This paper is organized as follows. In Sec.~\ref{se:naive}, we present the so-called \emph{naive} bound, which is the simplest  form of the inequalities. In Sec.~\ref{se:improved}, we show the gain obtained when removing the center-of-mass energy of the whole system and of the subsystems entering the inequality, with saturation for boson systems bound by harmonic forces: this corresponds to the \emph{improved} bound, also named Post bond.   In the case of systems with unequal masses or asymmetries in the potential, it is shown in Sec.~\ref{se:optimized} that the lower bound can be further bettered by writing a more flexible decomposition, at the expense of having to vary some parameters: this is named the \emph{optimized} bound, first developed for systems of three unequal masses, and further developed for larger systems. The methods developed for  unequal masses lead to some better results for bosons with both a pairwise interaction and a  potential from a fixed source.  The applications to the window for Borromean binding is reviewed in Sec.~\ref{se:coup:thr}.
The inequalities are applied to few-charge systems in Sec.~\ref{se:few-charges}, with rather deceiving  results, that illustrate the limitations of the method. In Sec.~\ref{se:mes-bar}, we list some inequalities among meson and baryon masses within  quark models, with either standard pairwise potentials with color factors, or the prescription of  a string of  minimal length. The case of tetraquarks is discussed in Sec.~\ref{se:tetra}, first in the approximation where the color wave function is  frozen, resulting a single-channel 4-body problem, and next when color mixing is accounted for. In the latter case, some technical developments of the Hall-Post formalism are required. The conclusions are presented in Sec.~\ref{se:conc}.

A word about the notation. It is difficult to carry exactly the same symbols for systems with a single mass and a single potential, and for systems with different masses with or without external attraction. For instance, the 2-body energy (unless specified, it is the ground-state) is denoted
$E_2(m)$ or $E_2(m_1,m_2)$ with $E_2(m_1,m_2)=E_2(2\, m_1\,m_2/(m_1+m_2))$, or 
$E_2(m;g)$ if the strength is specified (alternatively $E_2(m;V)$ or $E_2(m; g\,V)$)
and 
$E_2(m_1,m_2;g;g')$ in an external potential. Similarly for a 3-body system, we shall use
$E_3(m_1,m_2,m_3)=E_3(m_i)$, or 
$E_3(m_1,m_2,m_3;g_{23},g_{31},g_{12}) =E_3(m_i;g_{ij})$, which for symmetric systems, is abbreviated as
$E_3(m;g)$ or $E_3(m;V)$, and in case of an external field, we denote the energy as
$E_3(m_1,m_2,m_3;g_{23},g_{31},g_{12};g'_1,g'_2,g'_3) =E_3(m_i;g_{ij};g'_k)$.
\section{Naive bound}\label{se:naive}
\subsection{Naive bound for identical bosons}\label{subse:naive-bosons}
Consider three identical bosons of mass $m$ interacting through a pairwise, symmetric potential  $\sum V(r_{ij})$. Their Hamiltonian can be written as
\begin{equation}\label{eq:naive-decomp}
 H_3=\left[{\vis p 1\over 4\,m}+{\vis p 2\over 4\,m}+g\,V(r_{12})\right]+
 \left[{\vis p 2\over 4\,m}+{\vis p 3\over 4\,m}+g\,V(r_{23})\right]+
 \left[{\vis p 3\over 4\,m}+{\vis p 1\over 4\,m}+g\,V(r_{31})\right]~.
\end{equation}
Then the ground-state $E_3(m;g)$ is bounded by\footnote{If one aims at mathematical rigor, one cannot combine 2-body and 3-body Hamiltonians which act on different Hilbert spaces. A possible remedy consists in associating the 2-body term of, say, the pair $\{1,2\}$ with a wide harmonic well $\alpha(\vis p 3+ r_3^2)$, and take the limit $\alpha\to 0$.}
\begin{equation}\label{eq:naive}
 E_3(m;g)\ge E_\text{nai}=3\, E_2(2\,m;g)=\frac32\,E_2(m;2\,g)~,
\end{equation}
hereafter referred to as the \emph{naive} bound, 
where $E_2(m;g)$ is the ground-state of the 2-body system $(\vis p 1+ \vis p 2)/(2\,m)+ g\,V(r_{12})$ (which is assumed to exist; see Sec.~\ref{se:coup:thr} for a discussion on the coupling threshold, i.e., the minimal strength required to get a bound state). 

For instance, if $m=1$, and $V(r)=r^2$, one gets a lower bound $E_\text{nai}=9/\sqrt2\simeq6.364$ to be compared to $E_3=6\,\sqrt{3/2}\simeq7.348$.  In the linear case, $V(r)=r$, one gets $E_\text{nai}\simeq 5.567$ ($-E_2(1,1)$ is the first zero of the Airy function), vs.~$E_3\simeq6.132$ from an accurate numerical calculation \cite{1997CoPhC.106..157V}. In the gravitational case, $V(r)=-1/r$, the lower bound is $-3/2$, to be compared to $E_3\simeq -1.07$. 

The bound \eqref{eq:naive} is easily generalized to $N>3$. If one keeps $N'=2$, one gets
\begin{equation}\label{eq:N-naive}
 E_N(m;g)\ge \frac{N(N-1)}{2}\,E_2((N-1)\,m;g)~,
\end{equation}
and it is straightforward to extend to other values of $N'$.
\subsection{Naive bound for distinguishable particles}
We now consider the case of particles with different masses $m_i$ interacting with pairwise interactions $V(r_{ij})$, or $V_{ij}(r_{ij})$ if the potential differs from one pair to another. The immediate generalization of 
\eqref{eq:naive-decomp} reads
\begin{equation}
 H_3=\sum_{i<j}\left[ \frac{\vis p i}{4\,m_i}+\frac{\vis p j}{4\,m_j}+g_{ij}\,V_{ij}(r_{ij})\right]~,
\end{equation}
leading to
\begin{equation}\label{eq:naive:uneq}
 E_3(m_i;g_{ij})\ge E_\text{nai}=\sum_{i<j} E_2(2\,\mu_{ij};g_{ij}\,V_{ij})~, \quad \mu_{ij}=\frac{2\, m_i\,m_j}{m_i+m_j}~,
\end{equation}
where $\mu_{ij}$ is twice the reduced mass. 
For instance, for a system of masses $\{1,1,5\}$ interacting through a linear potential, the lower bound is $E_\text{nai}\simeq 4.986$, to be compared to the exact $E_3\simeq 5.457$. 
\subsection{Naive bounds for bosons in an external potential}\label{subse:bosons-ext}
\begin{figure}[ht!]
 \centering
\includegraphics[width=.45\textwidth]{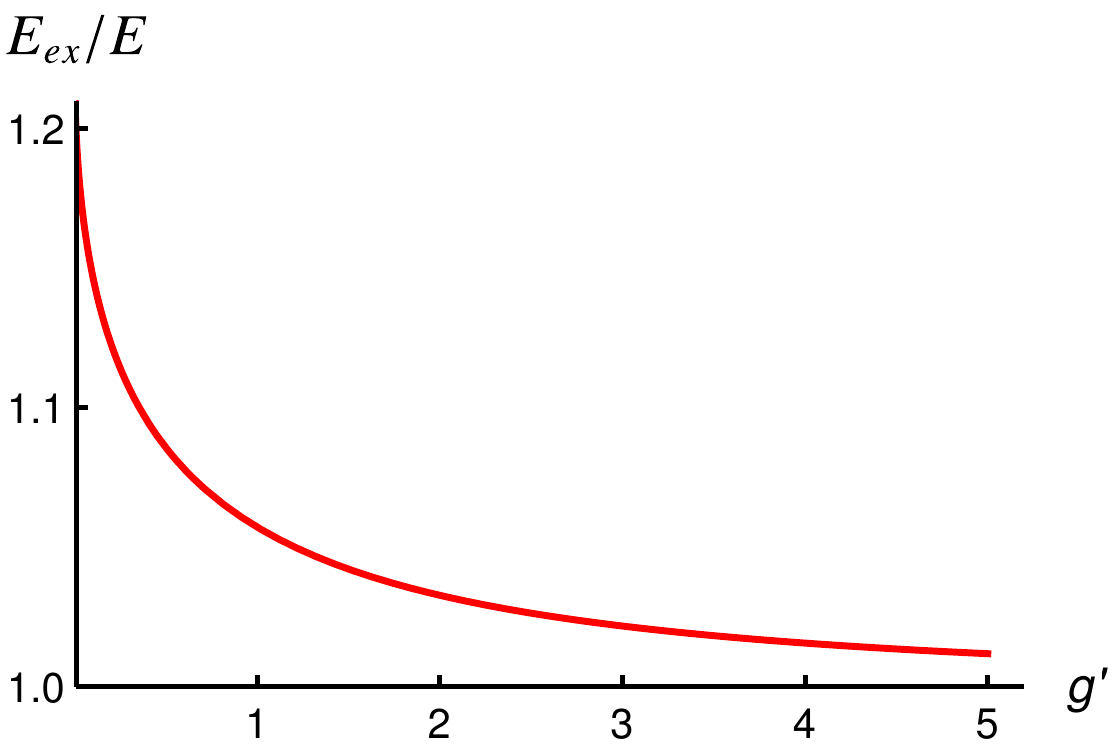}
 \label{fig:ext-pair-ho}
\end{figure}

There are many examples, for instance in the modeling of cold atoms, of systems with both a pairwise interaction and an external potential $U(r_i)$ acting on each particle. 
Let us consider the Hamiltonian 
\begin{equation}\label{eq:H-ext-pair}
 H_N(m;g;g')=\sum_i\left[\frac{\vis pi}{2\,m}+g'\,U(r_i)\right]+g\,\sum_{i<j} V(r_{ij})~,
\end{equation}
with ground state energy $E_N(m;g;g')$.  The simple decomposition
\begin{equation}
 H_N(m;g;g')=\sum _{i<j} H_2\left((N-1)\,m,g,g'/(N-1)\right)~,
\end{equation}
implies
\begin{equation}\label{eq:ext-pair-ho}
 E_N(m;g;g')\ge \frac{N\,(N-1)}{2}\,E_2\left((N-1)\,m;g;g'/(N-1)\right)=\frac{N}{2}\,E_2(m;(N-1)g;g')~.
\end{equation}
If $g'\gg g$, the inequality tends to be saturated, as the particles become independent and the energy $E_N(m;0;g')$ is proportional to $N$.   If $g'\to 0$, one recovers the case studied in Sec.~\ref{subse:naive-bosons}. If the confinement and the pairwise interaction are both harmonic, $H_N$ is exactly solvable: one can rescale to $m=g=1$, and study the bound as a function of $g'$. This is shown in Fig.~\ref{fig:ext-pair-ho}.
\subsection{Naive bound for fermions}\label{subse:naive-f}
In the case of three spinless fermions, the orbital wave function of the ground state is antisymmetric. For instance, for the harmonic oscillator $\sum_{i<j} V(r_{ij})=2\,(r_{12}^2+r_{23}^2+r_{31}^2)/3=\vec x^2+\vec y^2$, where $\vec x$ and $\vec y$ are the usual Jacobi coordinates, the wave function is $\Psi\propto \vec x\times \vec y\,\exp(-(x^2+y^2)/2)$. For another symmetric potential, $\Psi$ contains $\vec x\times \vec y$ times a more complicated symmetric function.  The 2-body subsystems are in a $p$-wave state (with a small admixture of  higher $\ell\ge3$ odd orbital momenta if the potential is not harmonic). So the bound \eqref{eq:naive} holds with $E_3$ being the lowest energy in the fully antisymmetric sector and $E_2$ the lowest energy with orbital momentum $\ell=1$.

For three fermions with spin and/or isospin, this is more delicate, as well as for $N\ge4$ particles. 
In this latter case, an astute trick has been devised. See, e.g., \cite{1969JMP....10..806L}.  The Hamiltonian is rewritten as 
\begin{equation}\label{eq:N-fermions}
 H_N=\sum_{i=1}^N h_i\left((N-1)\,m;g/2\right)=\sum_{i=1}^N\left[\sum_{j\neq i}\frac{\vis p j}{2\,(N-1)\,m}+\frac{g}{2}\,V(r_{ij})\right]~.
\end{equation}
where each $h_i(m,g)$ is a Hamiltonian describing $(N-1)$ independent particles of mass $m$ in the field $g\,V(r)$ of particle $i$. If the energy levels $\epsilon_n$ of the $h_i$ are known, as well as their degeneracy $g_n$, then one should make the counting of the occupied levels for $N-1$ particles, and calculate their cumulated energy in $h_i$. The exercise is done in \cite{1969JMP....10..806L} for the gravitational interaction $-G\,m^2/r$, with the results
\begin{equation}
 \begin{gathered}
 \epsilon_n((N-1)\,m;G\,m^2/2)=-\frac{(N-1)\, G^2 \,m^5}{4\,n^2}~,\qquad  g_n=n^2~,\\[4pt]
 \langle h_i((N-1)\,G\,m^2/2)\rangle \ge -\frac12{(N-1)^{4/3}\, G^2\,m^5}~,  \\[4pt]
 E_N(m;g)\ge  -\frac12{N\,(N-1)^{4/3}\, G^2\,m^5}~.
 \end{gathered}
 \end{equation}
 for spinless fermions, and suitable $2\,s+1$ factors for fermions of spin $s$. 
 
For bosons, the decomposition \eqref{eq:N-fermions} is equivalent to the naive bound, with the product of the effective mass and coupling being $(N-1)\,m\,g$ in each pair. We shall return to this identity when discussing the excited states, in Sec.~\ref{se:misc}.
\section{Improved bounds}\label{se:improved}
The naive bound never saturates the exact value of the ground state. The reason lies in the center-of-mass energy. The eigenvalue $E_3$ is the minimum of $H_3$ in the 3-body rest frame, but it is expressed in terms of the minimum of the subsystems. In the 3-body rest frame, the pair $\{1,2\}$ is not at rest, so omitting the overall kinetic energy of $\{1,2\}$ significantly underestimates the energy.  The remedy, already proposed in~\cite{0370-1298-69-12-409,0370-1328-90-2-309}, and independently rediscovered in \cite{1990NuPhB.343...60B}, consists of writing identities among intrinsic Hamiltonians. 
\subsection{Improved bound for three identical bosons}
Let $\tilde H_3=H_3-(\vec p_1+\vec p_2+\vec p_3)^2/(6\,m)$ and $\tilde H_2^{(ij)}(m)=(\vec p_i-\vec p_j)^2/(4\,m)+g\,V(r_{ij})$ denote the intrinsic parts of the 3-body and 2-body Hamiltonians. From the identity
\begin{equation}\label{eq:improved-dec}
 \tilde H_3(m;g)=\sum_{i<j} \tilde H_2^{(ij)}(3\,m/2;g)~,
\end{equation}
one gets the \emph{improved} bound
\begin{equation}\label{eq:improved}
 E_3(m;g)\ge E_\text{imp}=3\,E_2(3\,m/2;g)~,
\end{equation}
which is automatically better than the naive bound \eqref{eq:naive} since $E_2$ is a decreasing function of the reduced mass. For a harmonic oscillator, \eqref{eq:improved} becomes an identity. Thus it is the ultimate ``universal'' lower bound for the ground-state energy of three bosons in terms of 2-body energies.

For a linear interaction $V(r)=r$ and $m=1$, the bound $E_\text{imp}\simeq 6.1276$ is very close to the exact $E_3\simeq6.1323$. In the gravitational case, the bound $E_\text{imp}=-1.125$ approximates decently the exact $E_3\simeq -1.07$. 

\begin{figure}[ht!]
 \centerline{
 \includegraphics[width=.45\textwidth]{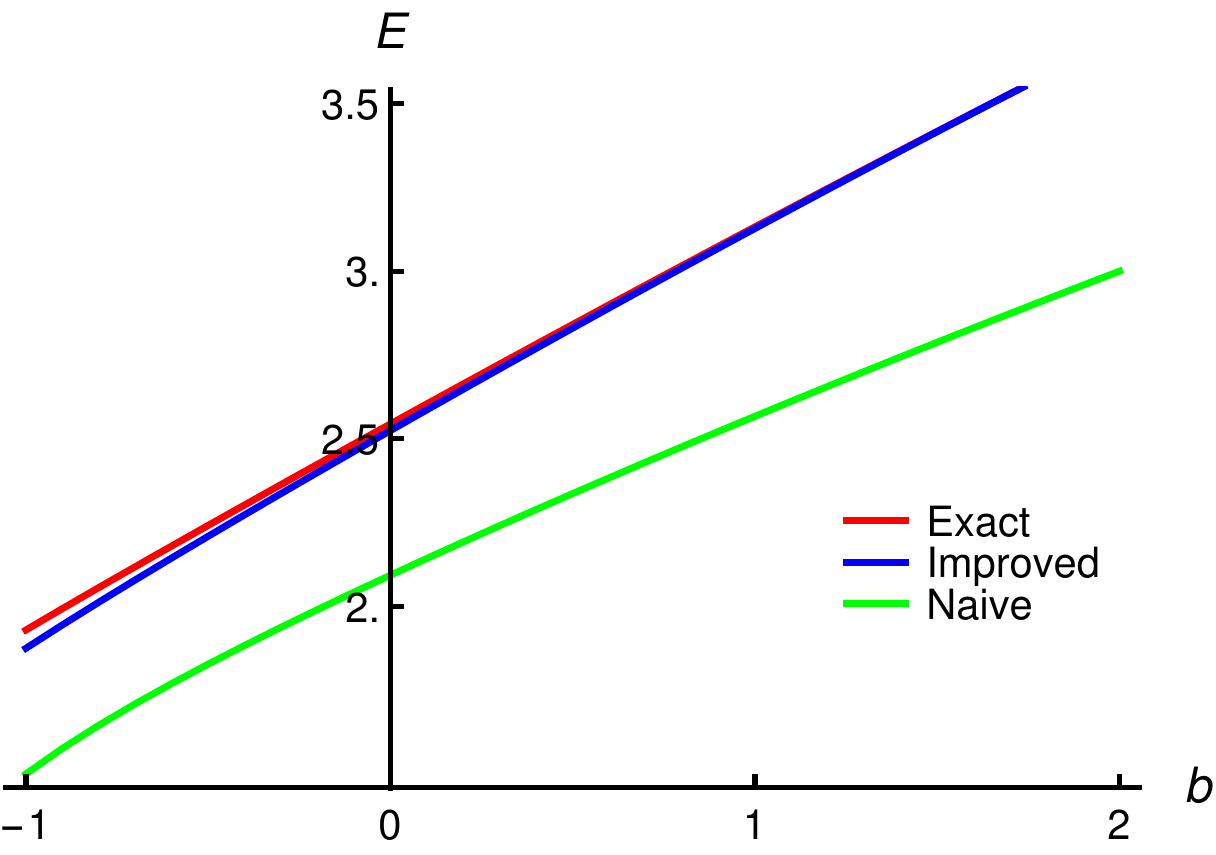}\quad \includegraphics[width=.45\textwidth]{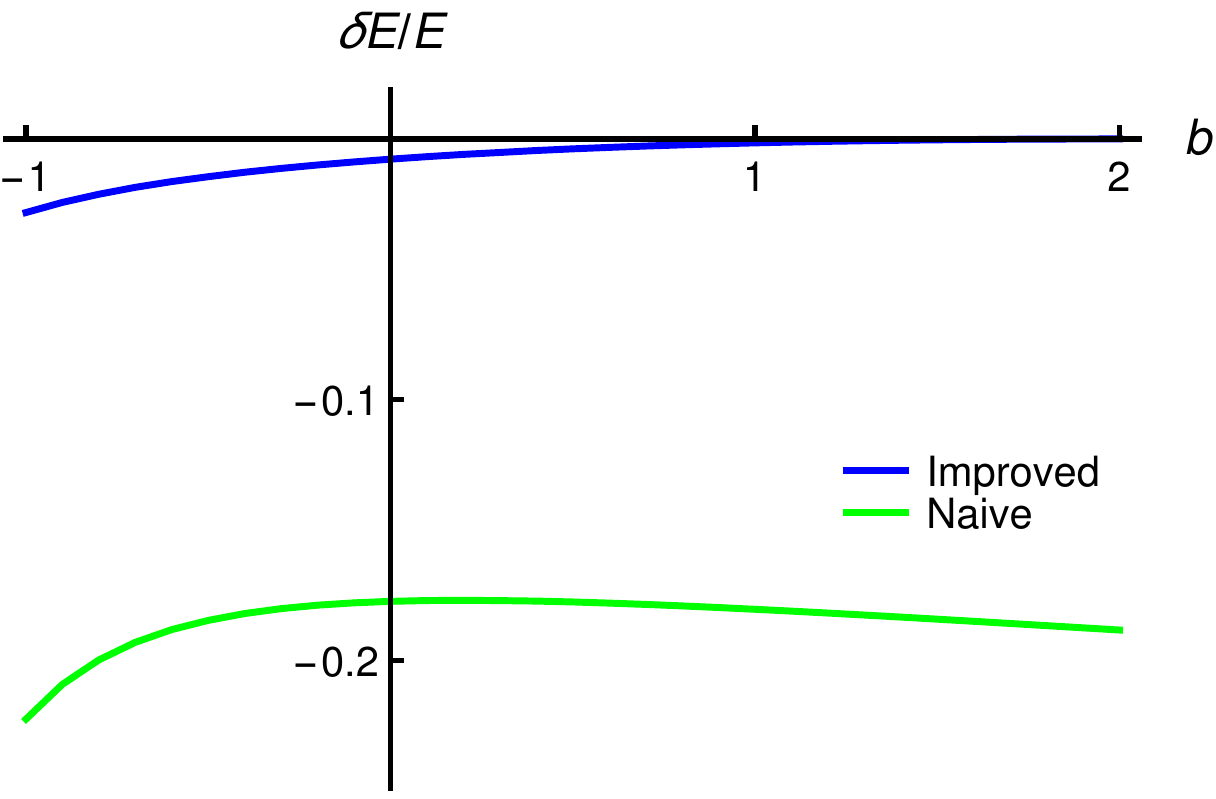}}
 \caption{Left: Comparison of the exact ground-state energy (red) for three bosons of unit mass in the  potential $V_b(r)=(r^b-1)/b$, and   the naive (green) or   improved bound (blue), as a function of the exponent $b$. Right: relative difference between the exact ground-state energy  and the bounds. }
 \label{fig:naive-imp-3}
\end{figure}

The improved bound is compared to the naive one and to the exact energy in Fig.~\ref{fig:naive-imp-3} for various power law potentials, more precisely for the family $V_b(r)=(r^b-1)/b$, which reduces to $V_0=\log r$ as $b\to 0$. There is a very slow deterioration of the accuracy when one departs from the harmonic-oscillator limit  $b=2$.  

\boldmath\subsection{Improved bound for $N$ identical bosons}\unboldmath
The generalization to $N>3$ identical bosons is straightforward. It relies on the identity
\begin{equation}\label{eq:dec:N-iden}
 \frac12\,\sum_{i=1}^N \vis p i-\frac{1}{2\,N}\,\left(\sum_{i=1}^N \vi p i\right)^{\!2}=\frac{2}{N}\sum_{i<j}\genfrac{(}{)}{}{0}{\vi p i - \vi p j}{2}^{\!2}~,
\end{equation}
and reads
\begin{equation}\label{eq:N-improved}
 E_N(m;g)\ge \frac{N(N-1)}{2} \, E_2(N\,m/2;g)=(N-1)\,E_2(m; g\,N/2)~.
\end{equation}
It is also saturated in the case of the harmonic oscillator. 

\begin{figure}[ht!]
 \centerline{
 \includegraphics[width=.45\textwidth]{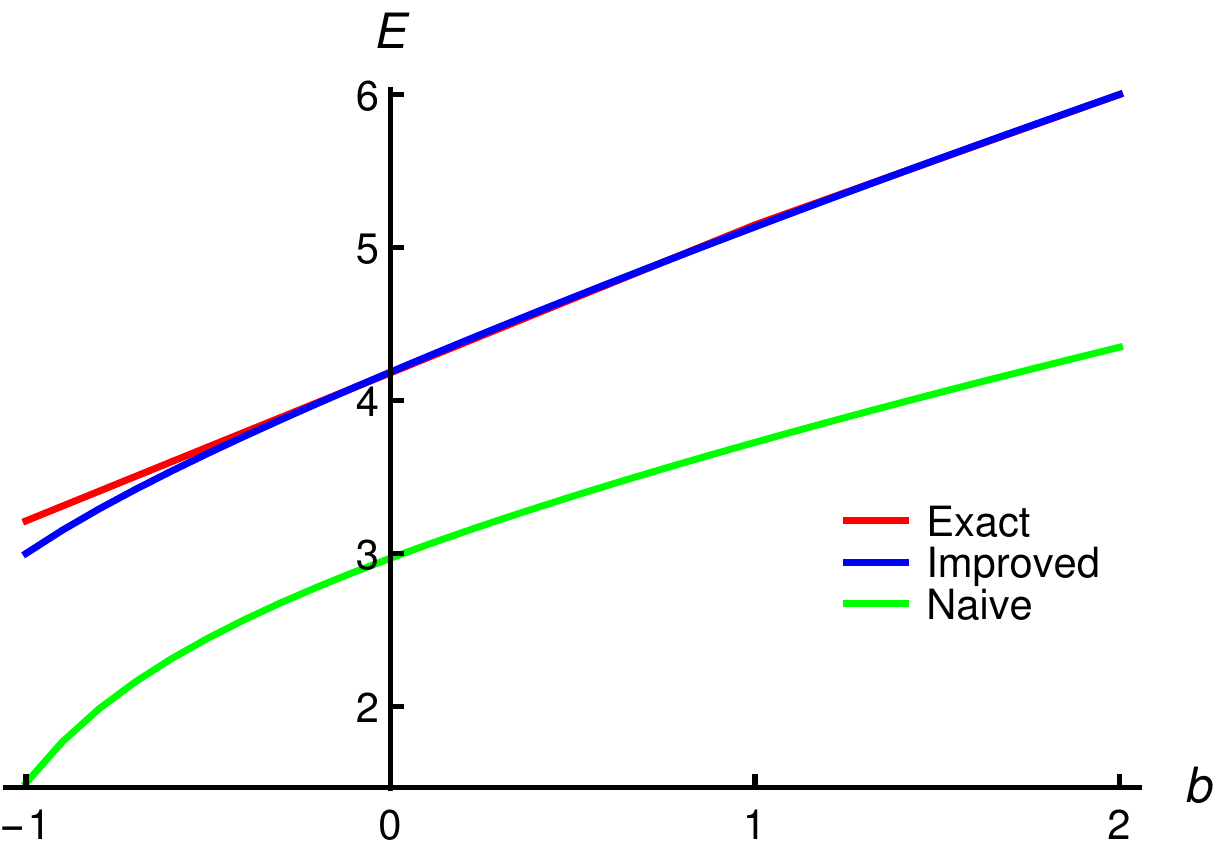}\quad \includegraphics[width=.45\textwidth]{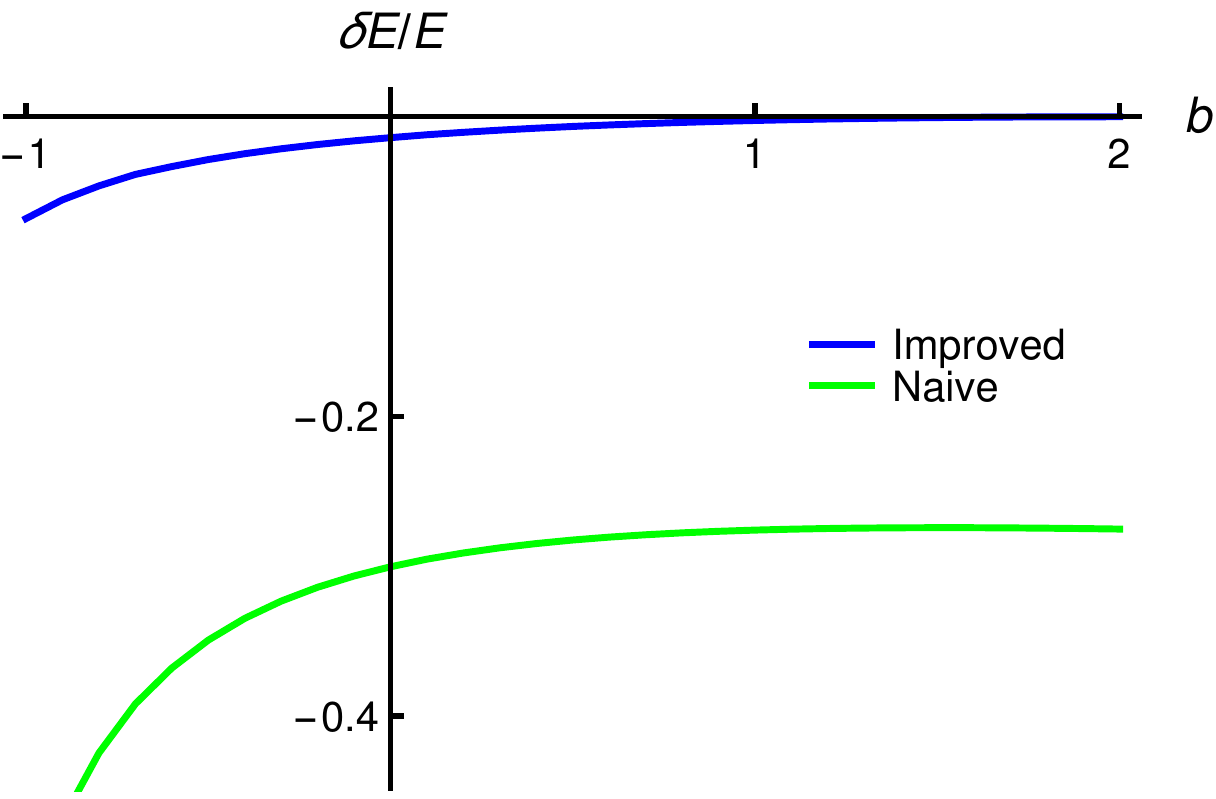}}
 \caption{Same as Fig.~\ref{fig:naive-imp-3}, but for four bosons.}
 \label{fig:naive-imp-4}
\end{figure}

The case of four-body systems is shown in Fig.~\ref{fig:naive-imp-4}. As expected, the naive bound is less accurate for the four-body system  than for the three-body one, since the overall kinetic energy of a pair is larger in the former case. 

\begin{figure}[ht!]
 \centering
 \includegraphics[width=.4\textwidth]{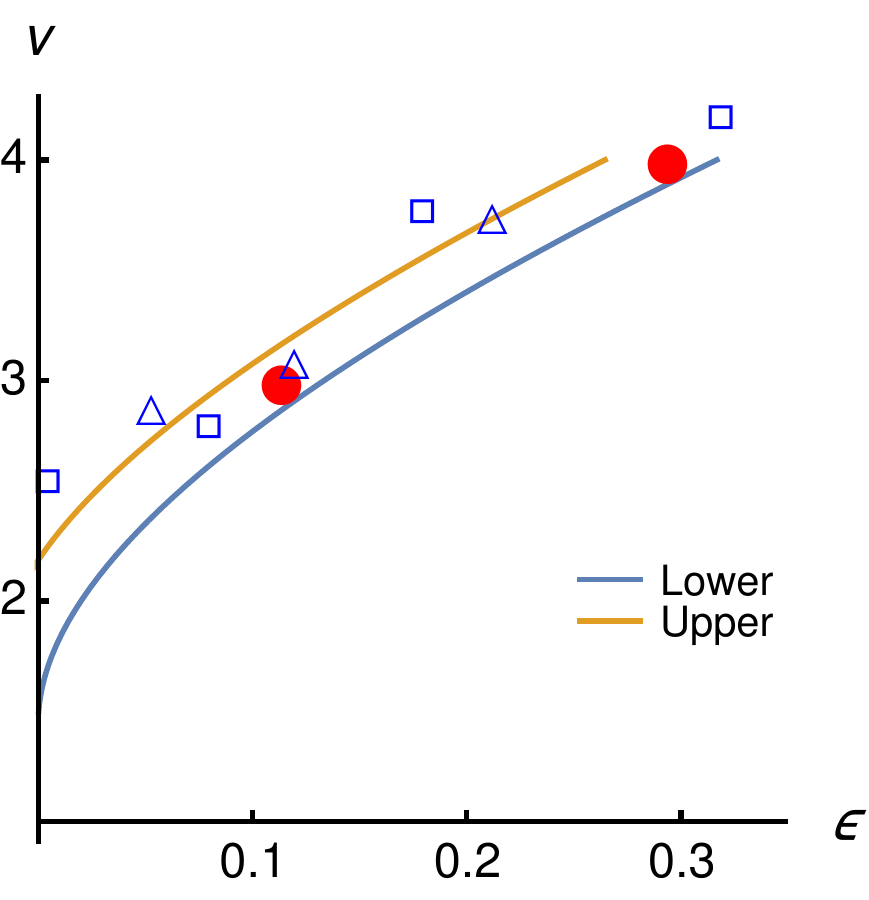}
 \caption{Upper variational bound and improved lower bound for $N$ bosons in the potential $-g\sum\exp(-r_{ij}/a)$. The plot displays $\epsilon=-E\,m\,a^2/(N-1)$ vs. $v=N\,m\,a^2\,g/2$. The red disk corresponds to two exact 4-body calculations, the blue squares (triangles)  to old quantum Monte-Carlo estimates by Kalos  \cite{1962PhRv..128.1791K} of 3-body (4-body) systems. This plot is copied from \cite{0370-1328-90-2-309}.}
 \label{fig:hpfig6}
\end{figure}

The power of \eqref{eq:N-improved} was illustrated by Hall and Post \cite{0370-1328-90-2-309} who used several potential shapes. We have redrawn in Fig.~\ref{fig:hpfig6} their Fig.~6, which includes some quantum Monte-Carlo estimates by \cite{1962PhRv..128.1791K} and two exact calculations of the 4-bosons system bound by an exponential potential. We use their notation
\begin{equation}
 \epsilon=\frac{|E_n|\,m\,a^2}{N-1}~,\qquad v=\frac{N\,m\,g\,a^2}{2}~,
\end{equation}
for the case of an exponential interaction $-g\,\sum \exp(-r_{ij}/a)$.  The upper curve corresponds to a variational approximation based on the simple Gaussian $\exp(-\alpha \sum r_{ij}^2)$.  Hall and Post stressed the existence of an almost universal curve for the binding energies when $E/(N-1)$ is plotted against $g/N$. Two exact energies corresponding to $N=4$ are shown to be just in between the lower and upper limits.  Some  3-body and 4-body energies  computed by Kalos \cite{1962PhRv..128.1791K} in some early quantum Monte-Carlo simulations have been shown by Hall and Post to violate the bounds. 

There have been several applications to systems of bosons with gravitational interaction $V(r)=-G\,m^2/r$, with the same \eqref{eq:N-improved} for the lower bounds, but different methods for calculating a variational upper bound. In \cite{1990NuPhB.343...60B}, a hyperscalar form is used for the trial wave function, $\Psi(r_{12}^2+\cdots)$, with the result, for large $N$
\begin{equation}
 -0.0625\,N^5\,G^2\,m^5< E_N< -0.0531\,N^5\,G^2\,m^5~.
\end{equation}
This upper bound is necessarily better than the one obtained from a Gaussian trial function $\exp[-\lambda(r_{12}^2+\cdots)]$, which is a particular case of hyperscalar function.  It has been noticed by Rebane \cite{2010TMP...162..347R} that the exponential $\exp(-G\,m^3 \sum_{i<j} r_{ij}/2)$ gives a slightly better $E_N< -0.0548\,N^5\,G^2\,m^5$. Presumably an exponential of the sum of distances, $\exp(-\alpha\,\sum_{i<j} r_{ij})$,  with an adjustable coefficient $\alpha$, or a generalized Feshbach-Rubinow approximation \cite{1955PhRv...98..188F}, $G(\sum_{i<j} r_{ij})$, will further improve this upper bound. 

\subsection{Improvements of the lower bounds for fermions}
\subsubsection{History}
The case of fermions is notoriously more difficult than for bosons. It turns out hard to get a better bound than the one of Sec.~\ref{subse:naive-f}, in which the $N$-fermion energy is expressed as $(N-1)$ two-body energies. 
There is some literature, but either with some controversy or a restricted domain of validity. 
Manning, for instance, suggested to modify the $s$-wave two-body energies entering the bound \cite{1978JPhA...11..855M}, but this resulted in a series that is barely convergent in the case of a harmonic oscillator and diverges for a linear interaction. 
The bound by Carr \cite{1978JPhA...11..291C} has been criticized by Balbutsev and Manning \cite{1978JPhA...11..291C,1978JPhA...11L.147B,1978JPhA...11L.143M}. 

A significant progress was achieved by  Hall \cite{1967PPS....91...16H,1995PhRvA..51.3499H}, improving previous attempts given in Ref.~\cite{1963RvMP...35..668C} and refs.\ there. It is based on a generalization of the the decomposition \eqref{eq:dec:N-iden}, which reads
\begin{equation}
 H=\sum_{i<j} \frac{2}{m\,N} \genfrac{(}{)}{}{}{\vec p_j-\vec p_i}{2}^{\!2} + V(r_{ij})~.
\end{equation}
Thanks to an optimal choice\footnote{The optimization of the Jacobi coordinates somehow anticipates the parametrization of the decomposition in the optimal bound for unequal masses, which is explained in Sec.~\ref{se:optimized}.}
of (non-orthogonal) Jacobi coordinates and associated relative momenta $\vec\pi_{ij}$, it is rewritten as
\begin{equation}
 H=\sum_{i<j} \left[\frac{\vec\pi_{ij}^2}{m\,N\,\lambda} + V(r_{ij})\right]~,
\end{equation}
with the best bound obtained for $\lambda=4/3$. It is 
\begin{equation}
 E_\text{H}=\frac{N}{2}\,\sum_{i=1}^{N-1} E_2^{i}(m\,N\,\lambda)~,
\end{equation}
as it is shown that the  requirement of antisymmetry imposes a summation over the $N-1$ first levels of $-\Delta/(m\,N\,\Lambda)+V$.  For the case of the harmonic oscillator, the ratio of the lower bound $E_\text{H}$ to the exact energy $E_\text{ex}$ can be calculated and compared to the one corresponding to the decomposition by L\'evy-Leblond et al.\  in Sec.~\ref{subse:naive-f},
\begin{equation}
 \frac{E_\text{H}}{E_\text{ex}}= \frac{\sqrt3}{2}\,\frac{N-1}{N-2}~,\qquad 
  \frac{E_\text{LL}}{E_\text{ex}}=\frac{\sqrt{N(N-1)}}{\sqrt{2}\,(N+1)}~. 
\end{equation}
The bound $E_\text{H}$ is better for $N>3$.

We now present below some more recent attempts. 
\subsubsection{Use of convexity inequalities}
Basdevant and Martin (BM) \cite{1996JMP....37.5916B} considered the class of power-law potentials $V_q=\sum r_{ij}^q$, and revisited convexity inequalities (that become identities for $q=2$) relating $V_q$\,, the sum of one-body potentials $U_q=\sum r_i^q$ and the term $W_q=|\sum \vi ri/N|^q$ acting on the center of mass. They derived new lower bounds  for the ground state energies of bosons and fermions.  In the latter case, this corresponds to a significant improvement with respect to the naive bound of Sec.~\ref{subse:naive-f}. More precisely, it is shown that
\begin{equation}\label{eq:convexity}
 \begin{aligned}
  2^{2-q}\,V_q+N^2\,W_q&\gtrless N\,U_q~\\
  V_q+N^q\,W_q&\lessgtr N\,U_q~,
 \end{aligned}
\end{equation}
with the upper inequality for $1\le q\le 2$, and the lower one for $q\ge2$, and a mere identity for $q=2$. When the kinetic energy is added, this is translated into the operator inequalities
\begin{equation}
\begin{aligned}
 \left[\sum_i\frac{\vis pi}{2\,m}+\sum _{i<j}  2^{2-q}\,r_{ij}^q\right]+\left[\frac{\vs P}{2\,N\,m}+N^2\,W_q\right]&\gtrless  \sum_i \left[\frac{\vis pi}{2\,m}+N\,r_i^q\right]~,\\
 \left[\sum_i\frac{\vis pi}{2\,m}+\sum _{i<j}  r_{ij}^q\right]+\left[\frac{\vs P}{2\,N\,m}+N^q\,W_q\right]&\lessgtr  \sum_i \left[\frac{\vis pi}{2\,m}+N\,r_i^q\right]~.
 \end{aligned}
\end{equation}

\begin{figure}[ht!]
 \centerline{
 \includegraphics[width=.4 \textwidth]{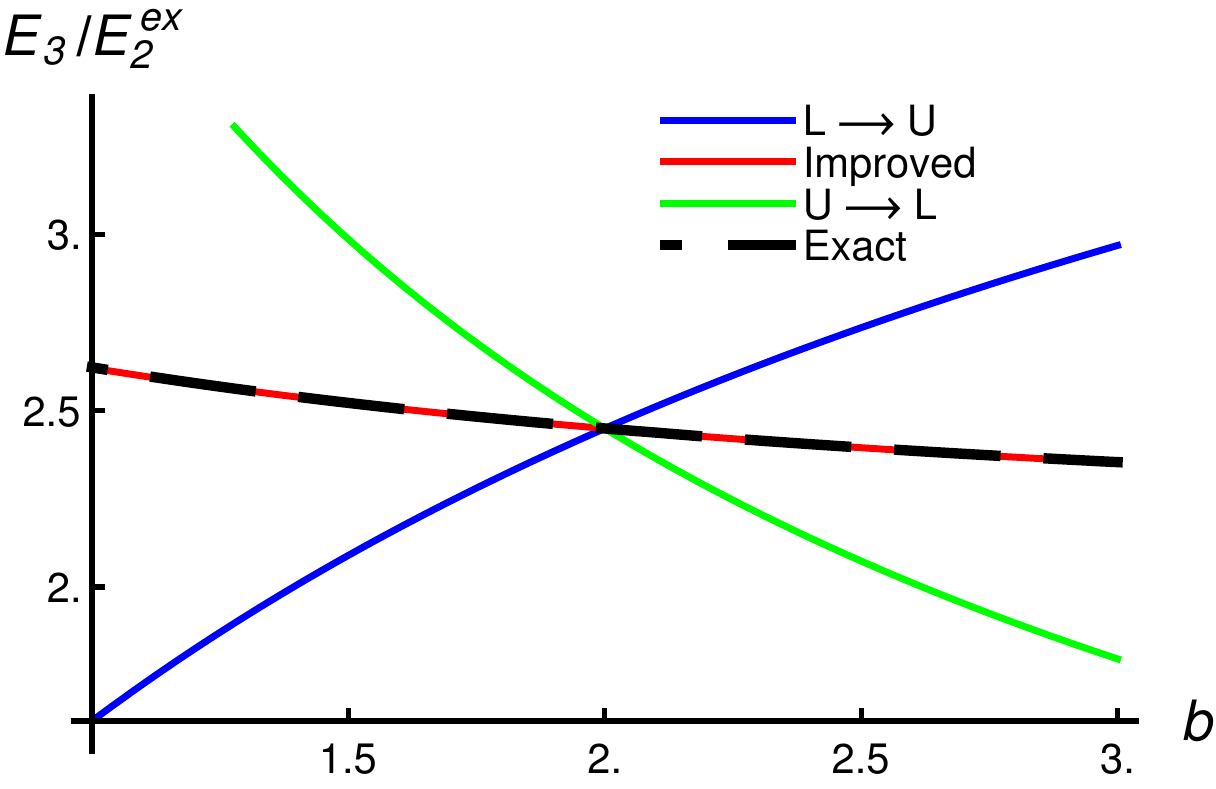}\hfill
  \includegraphics[width=.4 \textwidth]{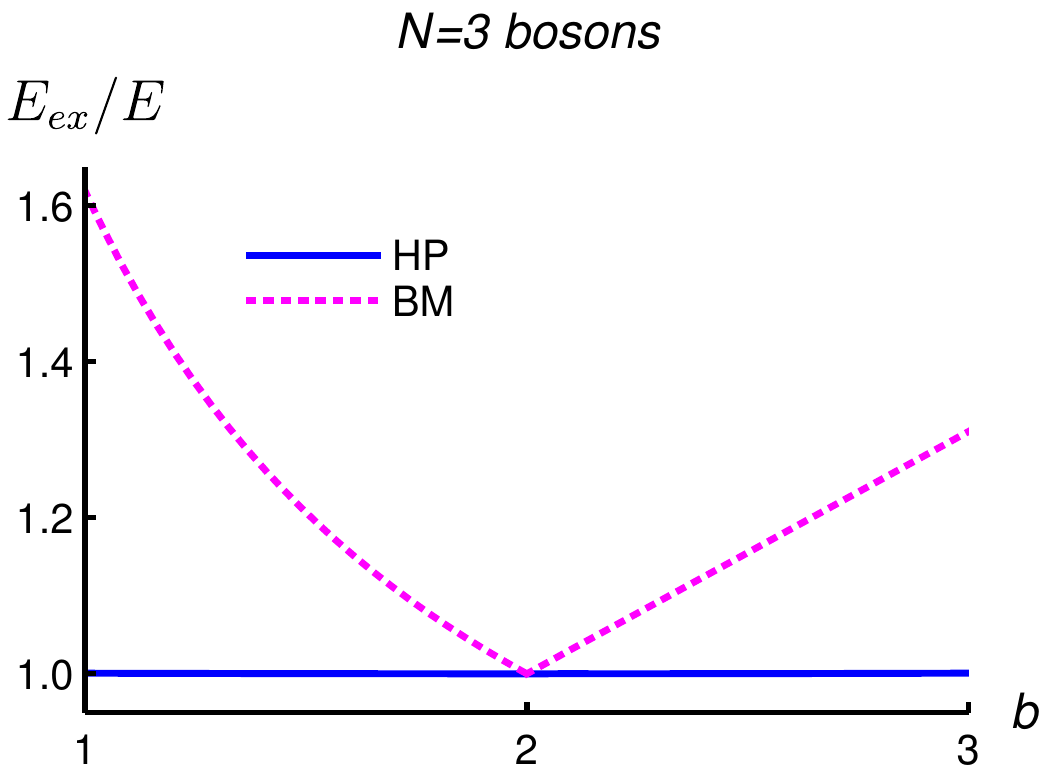}}
 \caption{Comparison of the HP  bound and the upper and lower bounds BM derived from the convexity inequalities for 3 bosons in a potential $r^b$, as a function of $b$.  Left: ratio of the 3-body energy to the exact 2-body energy (the improved HP bound, at this scale, is hardly distinguishable from the exact 3-body calculation). 
 Right: Ratio of the exact 3-body energy to the lower bound.}
 \label{fig:BM-3b}
\end{figure}

Let us consider the case of bosons, with the shortened notation  $\epsilon_q=E_2(1;1)$ to denote the ground-state energy of $\vec p^2+ |\vec r|^q$. For $1\le q\le 2$ (upper case) and $q\ge 2$ (lower case), respectively,  one gets
\begin{equation} 
\begin{aligned}
 2^{(4-2\,q)/(2+q)} \,E_N(q)+N^{4/(2+q)} \,2^{-q/(2+q)} \,\epsilon_q&\gtrless  N^{(2+2\,q)/(2+q)} 2^{-q/(2+q)}\, \epsilon_q~,\\
 E_N(q)+N^{2\,q/(2+q)} \,2^{-q/(2+q)} \,\epsilon_q&\lessgtr  N^{(2+2\,q)/(2+q)} 2^{-q/(2+q)}\, \epsilon_q~.
 \end{aligned}
\end{equation}
The bounds obtained for three bosons are shown in Fig.~\ref{fig:BM-3b}. This is clearly rather crude as compared to the improved bound \eqref{eq:N-improved}.
We conclude from this parenthesis dealing with bosons that the convexity inequalities \eqref{eq:convexity} quickly deteriorate as soon as one departs from the harmonic case $q=2$.  

\begin{figure}[ht!]
 \centerline{%
 \includegraphics[width=.45\textwidth]{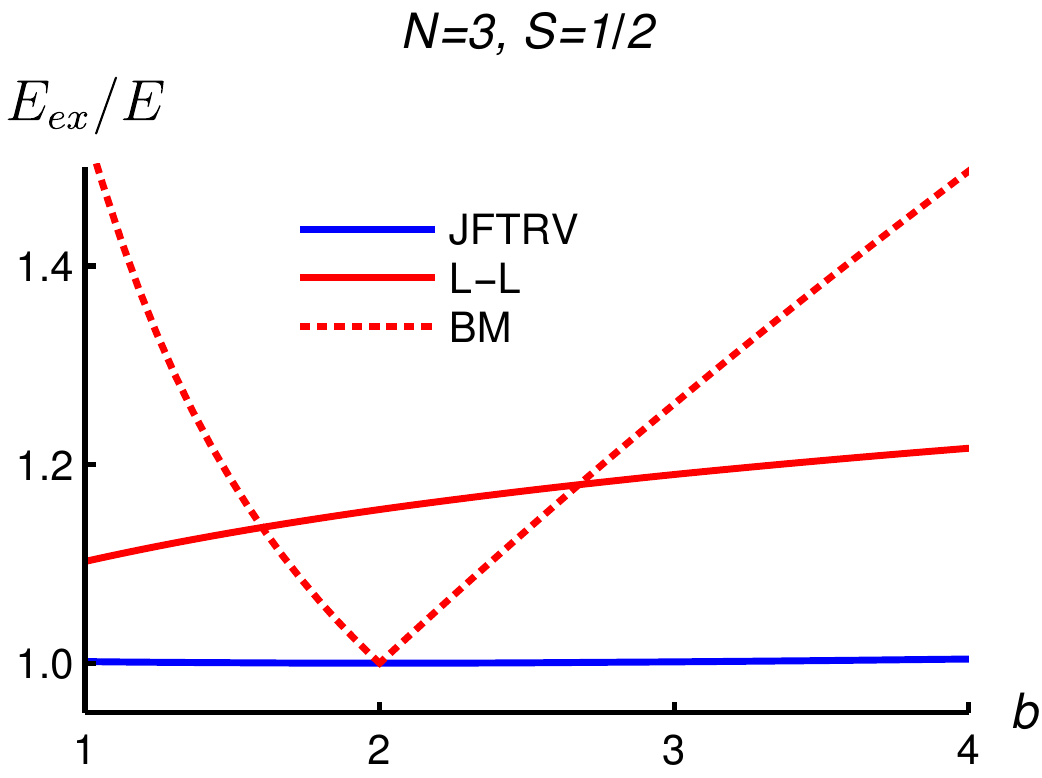}\hfill
  \includegraphics[width=.45\textwidth]{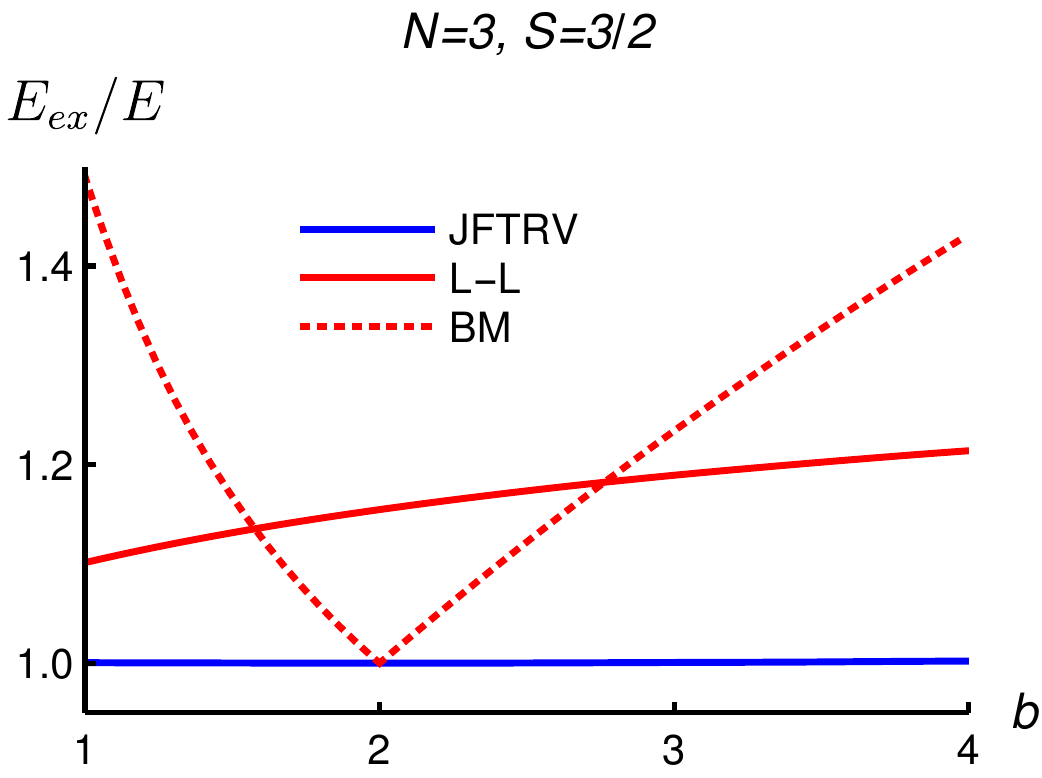}}
  \vskip .2cm
  \centerline{%
  \includegraphics[width=.45\textwidth]{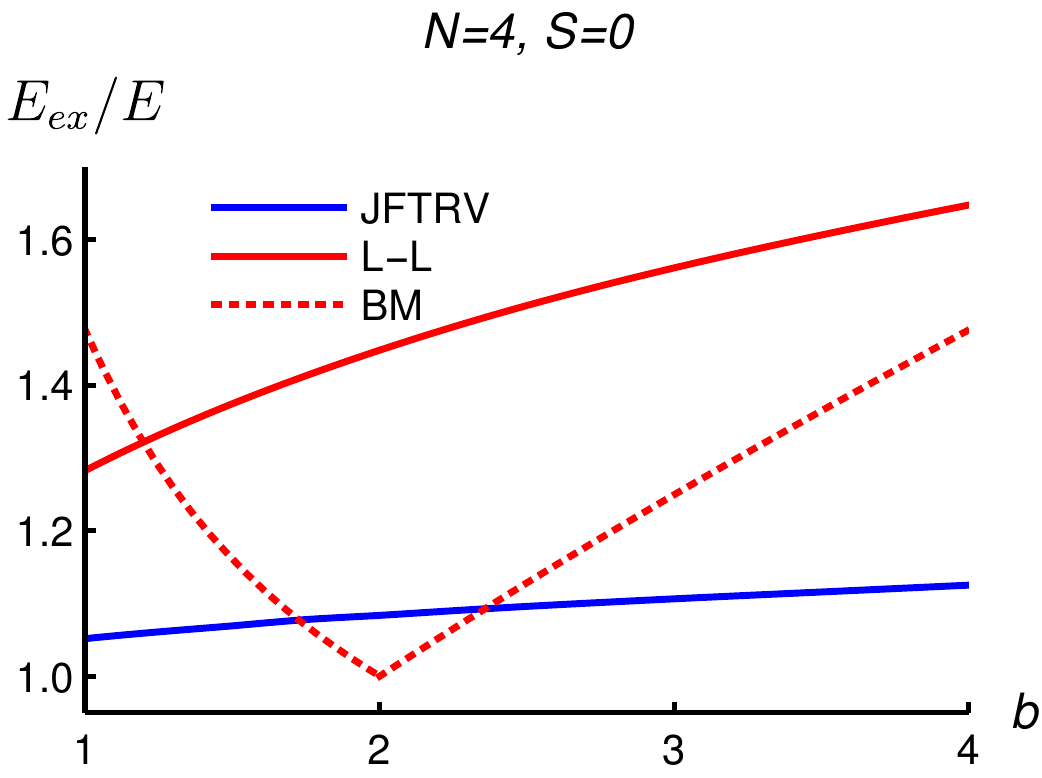}\hfill
    \includegraphics[width=.45\textwidth]{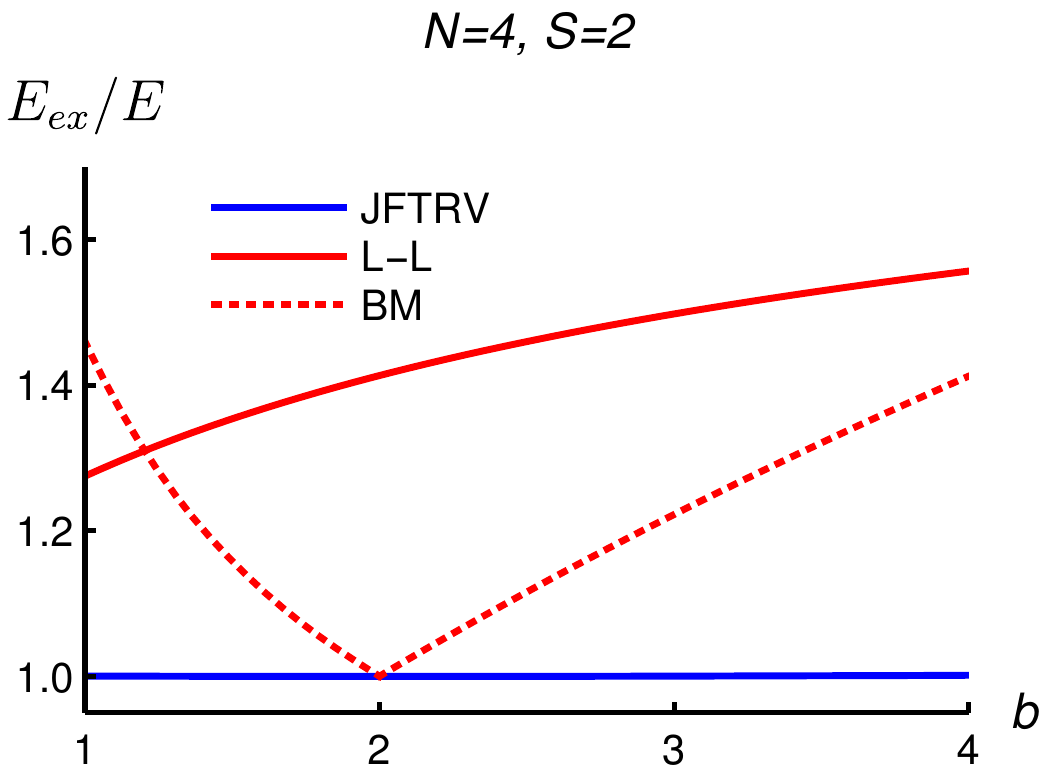}}
 \caption{Ratio of the exact energy to the BM lower bound,  the one of Juillet et al.\ \cite{2001PhRvB..63g3102J} (JFTRV) for $N=3$ or $4$ spin 1/2 fermions of total spin $S$, and the simple bound of L\'evy-Leblond \cite{1969JMP....10..806L} (L-L) discussed in Sec.~\ref{subse:naive-f}, for a potential $r^b$.}
 \label{fig:BM-F}
\end{figure}

We now come back to the case of fermions.  Let us denote $\epsilon_b(n,\ell)$ the single-particle energies of $\vec p^2+r^b$ with $b>1$ and $f_q(N)=\sum \epsilon_q(n,\ell)$ the cumulated  energies of $N$ fermions experiencing $\sum_i(\vis pi+ r_i^b)$. As in the textbooks of elementary chemistry, $f(N)=\epsilon(1S)+3\,\epsilon(1P)+\epsilon(2S)+\cdots$ for spinless fermions, and twice the previous expression for spin 1/2 fermions.  The ordering of $\epsilon(n,\ell)$ vs.\ $\epsilon(n',\ell')$ is discussed in \cite{Grosse:332762} for closest neighbors, but remains an issue in the general case, so that $f_q(N)$ has to be estimated empirically.  In Fig.~\ref{fig:BM-F} are shown the ratio of the exact energy to the BM bounds for three or four spin 1/2 fermions. There is a substantial improvement with respect to the naive bound  of Sec.~\ref{subse:naive-f}. 
\subsubsection{Group-theoretical considerations}
Juillet et al.\ \cite{2001PhRvB..63g3102J} adopted a different strategy. They studied the structure of a $N$-fermion wave function, more precisely how it can be constructed out of properly antisymmetrized $(N-1)$-body clusters. 

For instance, if $N$ fermions of spin 1/2 form a state of total spin $S=N/2$ (in absence of any other quantum number such as isospin, color, etc.), then the orbital function has to be antisymmetric, as well as the wave function of any subcluster. In particular
\begin{equation}\label{eq:fermionsNs2}
 E_N^{S=N/2}(m;g)\ge \frac{N}{N-2} E_{N-1}^{S=(N-1)/2}(N\,m/(N-1);g)~,
\end{equation}
which reduces to \eqref{eq:improved} for $N=3$.  For a more general value of the total spin $S$ (with $0\le S \le N/2$), it was shown in  \cite{2001PhRvB..63g3102J} that
\begin{multline} \label{eq:EvsN-1F}
E_N^S(m;g)\ge {N-1\over N(N-2)(2S+1)}\Biggl[ S(N+2S+2)\,E_{N-1}^{S-1/2}\left(m;{Ng\over N-1}\right) \\
{}+(S+1)(N-2S)\,E_{N-1}^{S+1/2}\left(m;{Ng\over N-1}\right)\Biggr]~.
\end{multline}
Some examples are shown in Fig.~\ref{fig:BM-F} for power-law potentials. 
%
%
\subsection{Further improvements}
Clearly, the previous HP bounds suffer from the fact that the expectation value of 2-body (or more generally $N'$-body with $N'<N$) subsystems within the exact $N$-body wave function is approximated by the ground state. 
Van Neck et al.~\cite{2001PhRvA..63f2107V}  have analyzed the corrections due to the excited states. Let us restrict here to the case of $N'=2$ subsystems.  Schematically, the $N$-body ground state is written as 
\begin{equation}
 \Psi_0^{(N)}\propto \sum_n \Psi_n^{(2)}(\vec x)\,\Phi_n(\vec y, \vec z, \ldots)~,
\end{equation}
where $\vec x=\vi r2-\vi r1$, $\vec y$, \dots is a set of Jacobi coordinates, besides the overall center of mass.  This results into replacing
\eqref{eq:N-improved} by 
\begin{equation}
 E_N(m;g)=\frac{N(N-1)}{2}\left[\alpha_0\,E_2^{(0)}(N\,m/2;g)+\alpha_1 \,E_2^{(1)}(N\,m/2;g)+\cdots\right]~,
\end{equation}
with $\sum \alpha_n=1$.  The usual HP bound comes from $E_2^{(n)}\ge E_2^{(0)}$. The first improvement reads
\begin{equation}
 E_N(m;g)\ge \frac{N(N-1)}{2}\left[\alpha_0\,E_2^{(0)}(N\,m/2;g)+(1-\alpha_0) \,E_2^{(1)}(N\,m/2;g)
 \right]~,
\end{equation}
the task is the evaluate the \emph{occupation numbers} $\alpha_0$ or an approximation that keeps the inequality. This is done in \cite{2001PhRvA..63f2107V}, together with some analogous developments for systems in a central field and for fermions. 

 In the case of three self-interacting bosons, it can be shown that the maximal occupation number for the ground state is the largest eigenvalue $\lambda$ of the integral equation
\begin{equation}\label{eq:73NDW}
\begin{gathered}
 (\lambda -1)\,G(r)= 2\,\int_0^\infty \mathrm{d}r'\,r'^2\, W(r,r')\, G(r')~,\\
W(r,r')=\frac12\int_{-1}^{+1} \mathrm{d}x\, g(|\vec r/2+\vec r'|)\,g(|\vec r'/2+\vec r|)~,
 \end{gathered}
\end{equation}
where $x=\hat{\vec r}.\hat{\vec r}{}'$,  $g(r)$ is the radial wave function (not reduced) of the ground state of the two-body problem, and $G$ its overlap with the 3-body wave function.   Equation \eqref{eq:73NDW} can be solved by discretization of  the integral (e.g., by Gauss-Hermite quadrature), resulting in a matrix equation whose unknown are the values $G(r_i)$ at the chosen points.  Some results of Table~I of \cite{2001PhRvA..63f2107V} are reproduced in Table~\ref{tab:NDW}. 
\begin{table}[!ht]
 \caption{ HP lower bound and its improvement HP$^*$ by seeking the optimal occupation number of the ground-state of the subsystems, for some power-law potentials $\sign(\beta)\,r^\beta$, where $\sign{\beta}=\beta/|\beta|$.}
 \label{tab:NDW}
 \centering 
 \vskip .1cm
  \begin{tabular}{rrrrr} \hline
  $\beta$ & \multicolumn{1}{c}{HP} &  \multicolumn{1}{c}{HP$^*$} &  \multicolumn{1}{c}{$E_3$} &  \multicolumn{1}{c}{$\lambda$} \\ 
  \hline 
$-1$       & $-1.1250$ &  $-1.1095$ &  $-1.0670$ & 2.9447 \\
1        & 6.1276  & 6.1309 & 6.1323 &  2.9978 \\
2        & 7.3485 & 7.3485 & 7.3485 &  3\phantom{.0000}\\
\hline
  \end{tabular}
\end{table}
Very likely, the method could be extended to the first excitations.\footnote{A correspondence with Dimitri van Neck and Michel Waroquier is gratefully acknowledged.}

\subsection{Improved bound for unequal masses}\label{subse:imp-uneq}
The  generalization of the improved bound to unequal masses is straightforward: the intrinsic 3-body Hamiltonian is written in terms of intrinsic 2-body Hamiltonians, where the momentum conjugate of $\vec r_j-\vec r_i$  is the usual combination $(m_i\,\vec p_j-m_j\,\vec p_i)/(m_i+m_j)$. It reads
\begin{equation}
 \tilde H_3(m_i;g_{ij})=\left[\frac{1}{\mu_{12}}\genfrac{(}{)}{}{0}{m_2\,\vi p1-m_1\,\vi p2}{m_2+m_3}^2+g_{12}\,V_{12}\right]+\text{circ.\ perm.}~,
\end{equation}
where
\begin{equation}\label{eq:eff-mass-imp-3}
\mu_{ij}=\frac{2\,m_i \,m_j\,(m_1+m_2+m_3)}{(m_i+m_j)^2}~,
\end{equation}
leading to 
\begin{equation}\label{eq:improved-uneq}
 E_3(m_i;g_{ij})\ge E_\text{imp}=\sum_{i,j} E_2(\mu_{ij};g_{ij})~,
\end{equation}
where $\{i,j,k\}$ is a direct permutation of $\{1,2,3\}$.
For instance, with $V(r)=r^2$ and masses $\{1,1,2\}$, one gets a lower bound $E_\text{imp}\simeq 6.621$, very close to the exact solution $E_3\simeq 6.674$, but saturation is not reached exactly. 

Needless to say that saturation is lost also for equal masses and unequal strengths. For instance for $m_i=1\ \forall i$ and an interaction $r_{12}^2+2\,r_{23}^2+3\,r_{31}^2$, the exact energy is $E_3\simeq10.28$, while the lower bound \eqref{eq:improved-uneq} is $E_\text{imp}=10.16$. For the linear $r_{12}+2\,r_{23}+3\,r_{31}$, the values are $E_3\simeq 9.655$ and $E_\text{imp}= 9.533$.

The result \eqref{eq:improved-uneq} is easily extended to more than three particles. For $N$ particles, the effective mass of the pair $\{1,2\}$ is 
\begin{equation}\label{eq:eff-mass-imp-N}
 \mu_{12}=\frac{2\,m_1 \,m_2(m_1+\cdots+m_N)}{(m_1+m_2)^2}~.
\end{equation}
\section{Optimized bound}\label{se:optimized}
The  improved bound of the previous section is excellent for bosons, and even becomes exact in the case of $V(r)\propto r^2$, but deteriorates for unequal masses (and/or unequal strengths). It is even observed, as in Figs.~\ref{fig:lin-NIOE} and \ref{fig:gra-NIOE} below,  that sometimes the naive bound provides a better result. The \emph{optimized} bound presented in this section is always better than the naive or the improved ones, and is saturated for the harmonic oscillator. However, it requires the  adjustment of  some parameters.

The early literature focused on the $(M,m,m,m,\ldots)$ configurations with $M$ infinite or $M\gg m$, which will be reviewed in Sec.~\ref{subse:central-field}. We first consider the most general distribution of masses. 
\subsection{Optimized bound for three-body systems}
The key is that many sub-Hamiltonians have the same 2-body spectrum with ground-state~$E_2$. The momentum $(m_1\,\vec p_2-m_2\,\vec p_1)/(m_1+m_2)$ is not compulsory as the conjugate of $\vec r_2-\vec r_1$, any combination $\alpha_2\,\vi p2-\alpha_1\,\vi p1$ can be used, provided $\alpha_1+\alpha_2=1$. Moreover, instead of subtracting from $H_3$ the kinetic energy of the center of mass, one can subtract a more general term proportional to the total momentum $\vec p_1+\vec p_2+\vec p_3$, whose expectation value vanishes within any eigenstate of $\tilde H_3$. This gives more flexibility.  The most general decomposition reads
\begin{equation}\label{eq:decom:opt3}
 H_3(m_i;g_{ij})=\left(\sum_{i=1}^3 \vi pi\right).\left(\sum_{i=1}^3 a_i\,\vi pi\right)
+\sum_{i<j}\left[x_{ij}^{-1}\genfrac{(}{)}{}{0}{u_{ij}\,\vec p_j-\vec p_i}{1+u_{ij}}^2+g_{ij}\,V_{ij}\right]~.
\end{equation}
The identification of the coefficients of $\vis p i$ and $\vec p_i.\vec p_j$ gives six relations, and one can express the parameters $a_i$ and $x_{ij}$ as  functions of the $u_{ij}$. Then one can maximize the lower bound
\begin{equation}
 E_\text{opt}=\max_{u_{12}, \ldots} \sum_{i<j} E_2(x_{ij}[u_{12}, \dots ];g_{ij})~.
\end{equation}
Note that the conditions for stationarity, which for each pair $\{k,\ell\}$ are of the form
\begin{equation}
 \sum_{i<j} \frac{\partial E_2(x_{ij},g_{ij})}{\partial x_{ij}}\, \frac{\partial x_{ij}}{\partial u_{k\ell}}=0~,
\end{equation}
with non-zero (actually negative) derivatives $E'_2$, imply that the $3\times 3$ determinant of the $\partial x_{ij}/\partial u_{k\ell}$ vanishes. This leads to
\begin{equation}\label{eq:condition}
 \prod_{k<\ell} u_{k\ell}=1~.
\end{equation}
This condition holds both for a symmetric potential $V_{ij}=V(r_{ij})$ with a unique function~$V$ or any asymmetric potential $V_{ij}=V_{ij}(r_{ij})$.

For instance, in the case of a linear interaction with masses $\{1,2,3\}$ and $V_{ij}=r_{ij}$, one reaches the maximum, $E_\text{opt}\simeq 5.144$ for $\{u_{k\ell}\}\simeq\{1.546,1.214,  0.533\}$. Conversely for equal masses $m=1$ and potentials $g_{ij}\,r_{ij}$ with $\{ g_{ij}\}=\{1,2,3\}$, one gets the best lower bound $E_\text{opt}\simeq 9.648$ for $\{u_{k\ell}\}\simeq\{0.732,\, 0.758,\,1.802\}$. In both cases, \eqref{eq:condition} is satisfied. 

For a quadratic interaction, the optimized bound coincides with the exact ground-state energy for unequal masses and/or strengths.

\begin{figure}[ht]
 \centering
 \includegraphics[width=.5\textwidth]{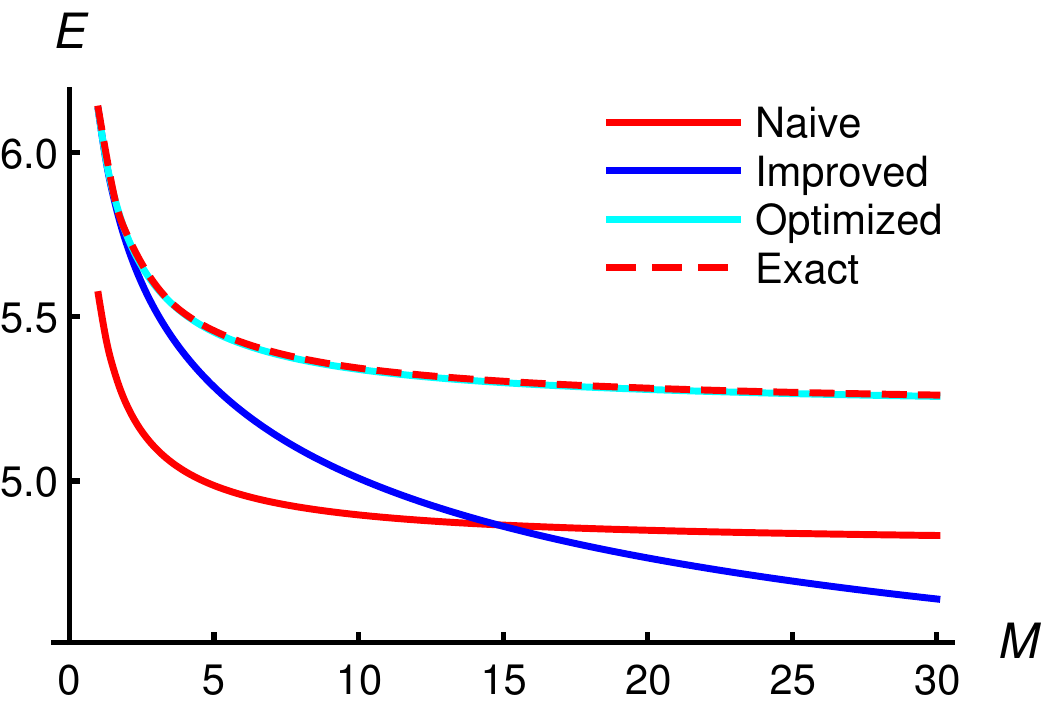}
 \caption{Comparison of the various lower bounds for a symmetric linear interaction, and masses $\{1,1,M\}$.  At the scale of the figure, the optimized bound is hardly distinguishable from the  upper bound obtained from the stochastic variational method of  Varga and Suzuki~\cite{1997CoPhC.106..157V}.}
 \label{fig:lin-NIOE}
\end{figure}

\begin{figure}[ht]
 \centering
 \includegraphics[width=.5\textwidth]{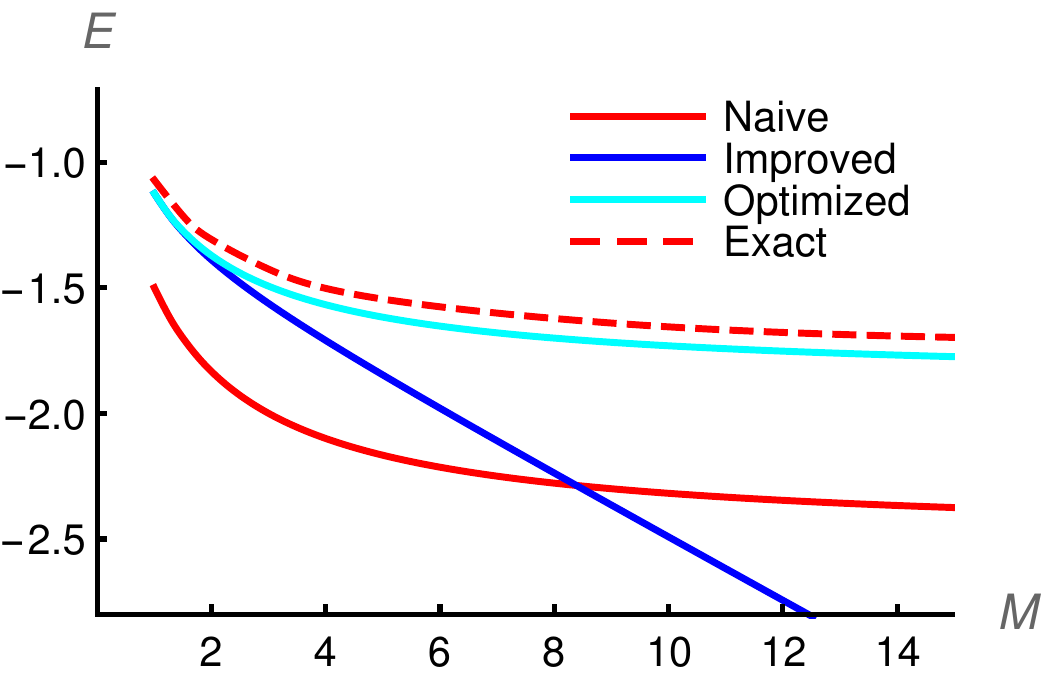}
 \caption{Comparison of the various lower bounds for a symmetric gravitational interaction, and masses $\{1,1,M\}$.  }
 \label{fig:gra-NIOE}
\end{figure}

As an illustration how this simple maximization helps, we show in Fig.~\ref{fig:lin-NIOE} the naive, improved and optimized lower bounds for a symmetric linear potential $r_{ij}$ and masses $\{1,1,M\}$, as a function of $M$.  The calculation is repeated in Fig.~\ref{fig:gra-NIOE} for a gravitational interaction, and the lower bounds become less accurate, as expected when one further departs from the harmonic case. 
\subsection{Optimized bound for four-body and larger systems}\label{subse:opt4}
When going from the improved to the optimized bound for three-body systems, we had \textsl{i)} to vary some parameters, \textsl{ii)} to subtract an operator more general than the overall kinetic energy, and  \textsl{iii)} to introduce  in the 2-body subsystems a more general combination of individual momenta, namely $\alpha_j\,\vec p_j-\alpha_i\,\vec p_i$, with $\alpha_i+\alpha_j=1$.

When deriving an optimized lower bound for four-body systems, the third of the above extensions should be pushed further. Namely, the conjugate of  $\vec r_j-\vec r_i$ is not  restricted to be a combination of $\vec p_i$ and $\vec p_j$. 
Twice the momentum conjugate to $\vec r_j-\vec r_i$ is written as $\sum_k x_{ij,k}\,\vi p k$. But as one may add any vector proportional to the total momentum $\sum_k \vi p k$ and has to impose the normalization of the commutators of the position and momentum variables, one can choose $x_{ij,i}=1$ and $x_{ij,i}=-1$.
Eventually, the decomposition reads
%
\begin{multline}\label{eq:opt-4masses}
 H_4=\sum_{i=1}^4\frac{\vis p i}{2\, m_i}+\sum_{i<j} V_{ij}= \left(\sum_{i=1}^4 \vi p i\right).\left(\sum_{i=1}^4 b_i\, \vi p i\right)\\
{}+ \sum_{i<j}\left[ \frac{a_{ij}^{-1}}{4} \left({
\sum_{k=1}^4 x_{ij,k}\,\vi p k}\right)^2 + V_{ij}\right]~.
\end{multline}
%
The identification of the coefficients gives 10 equations, from which one can determine the reduced  masses $a_{ij}$ and the auxiliary quantities  $b_i$ as functions of the parameters $x_{ij,k}$ with $k\neq i$ and $k\neq j$. As for the 3-body case, when the cumulated energy $S=\sum_{i<j} E_2(a_{ij})$ is maximized, the twelve conditions $\partial S/\partial x_{ij,k}=0$ are not independent. Seven general relations can be written down, and for each set of masses and potentials, the lower bound $S$ is optimized  by varying only five parameters. Details are given in~\cite{1998FBS....24...39B}. The study has been pushed further by Zouzou et al. \cite{2006JPhA...39.7383E,2006JPhA...39.5857E,2009FBS....46..199B}, for more than four particles.

\begin{figure}[ht!]
 \centering
 \includegraphics[width=.5\textwidth]{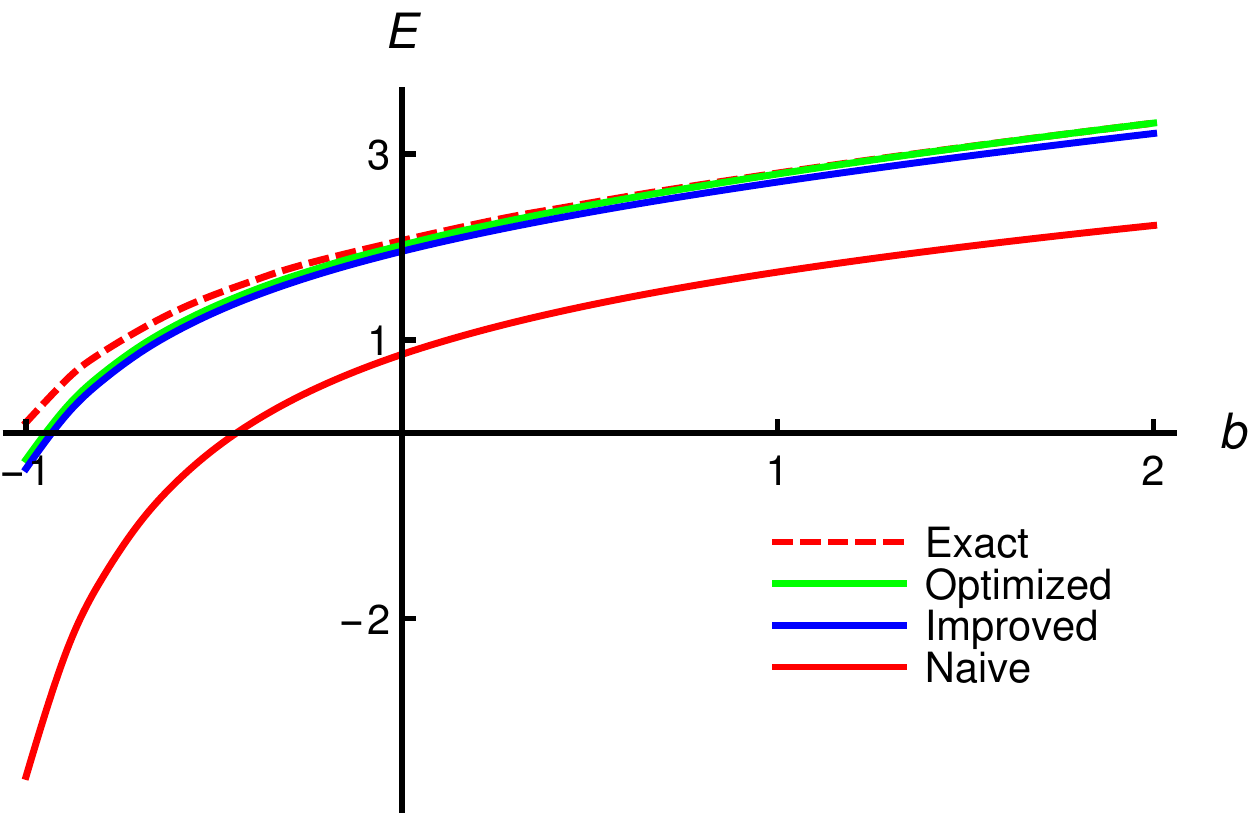}
 \caption{Comparison of various Hall-Post bounds to the exact energies for masses $\{1,2,3,4\}$ in the potential $V(r)=(r^b-1)/b$, as a function of the exponent $b$.}
 \label{fig:1234}
\end{figure}

As a first illustration, we consider a set of  masses $\{m_i\}=\{1,2,3,4\}$ experiencing a power-law interaction  $V_b(r)=(r^b-1)/b$.
The comparison of the exact energy and the naive, improved and optimized bounds is done in Fig.~\ref{fig:1234}.

\begin{figure}[ht!]
 \centering
 \includegraphics[width=.5\textwidth]{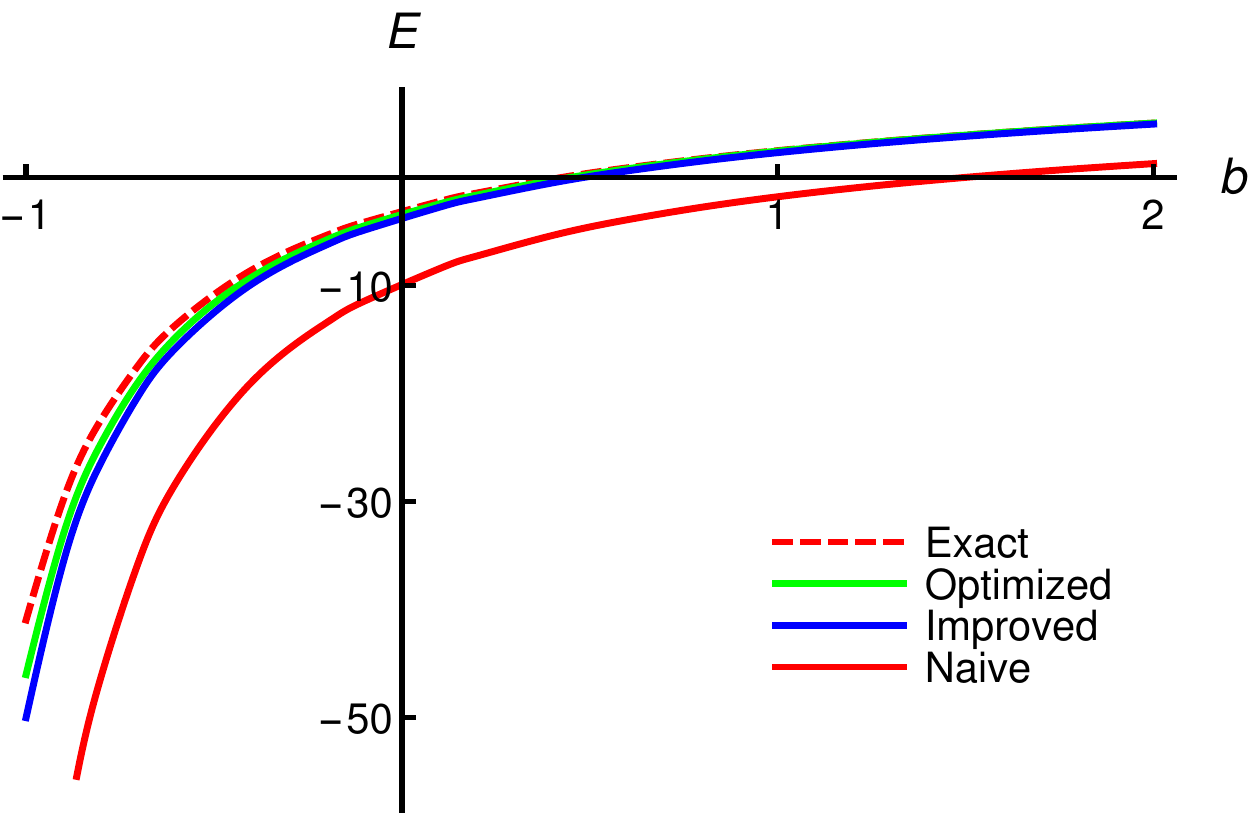}
 \caption{Comparison of various Hall-Post bounds to the exact energies for equal masses $m_i=1$ in the potential $\sum g_{ij}\,(r_{ij}^b-1)/b$, with $g_{ij}=i+j$, as a function of the exponent $b$.}
 \label{fig:ij}
\end{figure}

Instead of introducing different masses in a symmetric potential, one can consider a potential with unequal strength factors among the pairs, experienced by equal-mass particles. In Fig.~\ref{fig:ij} is shown the case of equal masses $m_i=1$ experiencing the same shifted power-law potential as above, but with a multiplying factor $g_{ij}=i+j$ for each pair. Again, the optimized bound is slightly better, and matches the exact energy for harmonic confinement.


One can of course combine the asymmetries in the potential and in the masses. With  masses $m_i=i$ and the above $g_{ij}=i+j$, the exact HO energy $E_4\simeq 18.17$ is recovered by adjusting properly the parameters of the optimized bound. 
\subsection{Optimized bound for four-body systems in terms of three-body subsystems}\label{subse:4outof3}
There are circumstances where the 4-body Hamiltonian  is under control, but where some of its 2-body subsystems, if taken separately, either do not bind (see Sec.~\ref{se:few-charges}) or are not bounded below (see Sec. \ref{sub:M-color}).  The former case corresponds to the repulsion between identical charges, the latter one to  a linear confinement $\sum g_{ij} r_{ij}$ with $g_{12}<0$ but the other $g_{ij}$ being sufficiently positive to ensure that  the whole potential remains always positive, thanks to the triangular inequalities.  Then a lower bound in terms of 2-body energies is meaningless, but one can get an interesting Hall-Post bound from 3-body subsystems, at least for a certain range of values for the couplings. 

For simplicity, and in view of the applications to hydrogen-like molecules $(M^+,M^+,\linebreak[0]{}m^-,m^-)$ and tetraquarks $(QQ\bar q\bar q)$, we shall restrict ourselves to  study the systems $(M,M,\linebreak[0]{}m,m)$ with at most two values for the  masses and two  couplings, $g=g_{12}=g_{34}$ and $g'=g_{13}=g_{14}=g_{23}=g_{24}$. For the harmonic oscillator, the 2-body systems exist for $g>0$ and $g'>0$; the 3-body ones for $g'>0$ and $2\,g+g'>0$, while the 4-body system requires only $g'>0$ and  $g+g'>0$. 
We shall concentrate on the cases where binding exists for both the 4-body system and its 3-body subsystems, and derive the corresponding Hall-Post inequalities. To our knowledge, these details have never been spelled out in the literature. 

For equal masses $m$, there is a straightforward improved bound, as previously defined. One can match the intrinsic part $\tilde H_4$ of the 4-body Hamiltonian with a sum of intrinsic 3-body Hamiltonians $\sum_i \tilde{H}_3^{(i)}(m')$, where $i$ denotes the missing particle and $m'=8\,m/3$. It reads
\begin{multline} \label{eq:E4from3-1}
 \sum_{i=1}^4 \frac{\vec p_i^2}{2\,m}-\frac{(\vi p 1+\cdots+\vi p 4)^2}{8\,m}+\sum_{1\le i<j} ^{j=4} V_{ij}\\
{} =\left[ \frac{3 (\vec p_1^2+\vec p_2^2+\vec p_3^2)}{16\,m} -\frac{(\vec p_1+\vec p_2+\vec p_3)^2}{16\,m}+\frac{V_{12}+V_{23}+V_{31}}{2}\right]+\cdots
\end{multline}
 This corresponds to a lower bound
\begin{equation}
 \label{eq:E4from3-2}
 E_4(m;V)\ge E_\text{imp}=4\,E_3(8\,m/3;V/2)~.
 \end{equation}
 For instance, with a harmonic interaction $\sum_{i<j} r_{ij}^2$ for the 4-body problem, and mass $m=1$, the lower bound saturates the exact value $E_4=9\,\sqrt2$. 
 With a purely linear $V(r)=r$ or gravitational $V(r)=-1/r$, one gets, respectively, $E_4\simeq 11.148$, $E_\text{imp}\simeq11.143$ and $E_4\simeq-2.788$, $E_\text{imp}\simeq-2.857$. 
 
 In the case of unequal masses, $(M,M,m,m)$, the simplest decomposition does not modify the sharing of strengths  within the subsystems. It corresponds to $u=0$ in:
 \begin{multline}\label{eq:E4from3-3}
 \frac{\vis p 1+\vis p 2}{2\,M}  + \frac{\vis p 3+\vis p 4}{2\,m}-\left( \sum_{i=1}^4 \vi p i \right).\left(A\,\vi p 1+A\,\vi p 2+ a\,\vi p 3+a\,\vi p 4\right)+\sum_{1\le i<j} ^{j=4} V_{ij}\\
 \begin{aligned}
 &{}= \left [\frac{\vis p 1+\vis p 2}{2\,x_1}+\frac{\vis p 3}{2\,x_2}-\frac{\vi p 1+\vi p 2+\vi p 3}{2\,x_1+x_2}+\frac12\left(V_{12}+(1+u)(V_{13}+V_{23})\right)\right] + \{ 3 \leftrightarrow 4\}\\
&{}+  \left[\frac{\vis p 3+\vis p 4}{2\,x_3}+\frac{\vis p 1}{2\,x_4}-\frac{\vi p 3+\vi p 4+\vi p 1}{2\,x_3+x_4}+\frac12\left(V_{34}+(1-u)(V_{13}+V_{14})\right)\right] + \{1 \leftrightarrow 2\}~. 
\end{aligned}
 \end{multline}
 
 Consider first the simplest choice $u=0$.
The identification gives four relations among  the parameters $A$ and $a$ and the four masses $x_i$, and one can calculate $x_3$ and $x_4$ from $x_1$ and $x_2$. By varying the latter, the lower bound $2\,E_3(x_1,x_1,x_2)+2\,E_3(x_3,x_3,x_4)$ can be optimized and approaches closely the exact energy, for instance 10.902 vs.\ 10.917 for $m=1$ and $M=2$ in the harmonic oscillator. 

An improvement consists of introducing some flexibility in the sharing of the $(M,m)$ interaction terms, namely $u\neq0$ in \eqref{eq:E4from3-3}. Varying $u$ as well and the masses $x_1$ and $x_2$ leads to exact saturation in the case of the harmonic oscillator. 

The same decomposition works for unequal strengths associated with equal masses, or a combination of unequal masses and unequal strengths. For instance, for $\sum_{i<j} g_{ij}\,r_{ij}^2$ and  $g_{12}=g_{34}=1$ and other $g_{ij}=2$, one gets saturation of the exact energy $E_4=16.39$ with masses $x_1=x_3=3$ and $x_2=x_4=2.2$ in the 3-body clusters, and a balanced sharing of strengths ($u=0$).  If the masses are changed to $M=2$ and $m=1$, saturation of the exact energy can still be obtained, at $E_4\simeq 14.07$ for $x_i\simeq\{7.2, 3.6, 2.6, 2.9\}$ and $u=-1/3$.  One could check, however, that a function is stationary near its maximum. The improved bound, with a mere subtraction of the center-of-mass energy, and corresponding to the masses
\begin{equation}\label{eq:masses-imp}
 \{x_i\}=\left\{\frac{4 M (m+M)}{m+2 M},\frac{4 m (m+M)}{m+2 M},\frac{4 m (m+M)}{2 m+M},\frac{4 M (m+M)}{2 m+M}\right\}~,
\end{equation}
and $u=0$, is just about 0.5\% below the optimized maximum.  More examples will be given when discussing hydrogen-like molecules and tetraquarks.

At the end of this section, we insist  that the decomposition of 4-body in terms of 3-body clusters is useful mostly when some 2-body subsystems do not bind or are ill defined, for instance with an anti-confining interaction. In case of very unequal masses, or strength coefficients close to the domain of collapse for 3-body systems, the optimization of the lower bound becomes delicate, with the  mass parameters at the edge of the allowed domain.
\subsection{Application to bosons in a central field}\label{subse:central-field}
The methods developed for unequal masses lead to a significant progress for the lower bound to a system of bosons experiencing both an external interaction and pairwise forces.  The Hamiltonian is given by  \eqref{eq:H-ext-pair} of Sec.~\ref{subse:bosons-ext}, where the naive form of the lower bound was given. 

For such systems, there is no improved bound in the sense that it is in general impossible to separate explicitly the center-of-mass motion, an exception being the case of a harmonic interaction. Calogero and Marchioro \cite{1969JMP....10..562C} have proposed various decompositions, which in our notation read
\begin{equation}
 H=H_1+H_2=
\sum_i\left[\frac{(1-x)\,\vis pi}{2\,m}+g' \,U(r_i)\right] + \left[
\sum_i\frac{x\,\vis pi}{2\,m}+g\,\sum_{i<j} V(r_{ij})\right]~,
\end{equation}
and are to be optimized by varying the parameter $x$ that governs the sharing of  the kinetic energy. Then the improved bound can be applied to the self-interacting system $H_2$.  For three-bosons with $m=g=g'=1$, and potentials $U(r)=r^2$ and $V(r)=r$, one obtains $E\ge9.8236$ if $H_2$ is treated exactly, not too far from the exact value (obtained numerically), which is about 9.8346. For comparison, the naive bound of Sec.~\ref{subse:bosons-ext} gives $E_\text{nai}\simeq 9.663$, while the above bound by Calogero and Marchioro, once optimized on $x$, gives a rather poor 9.2914. 

Now, we can consider the above system as a $(N+1)$-body system with masses $(M,m,m,\ldots)$ in the limit where $M\to \infty$, and search for the optimized bound. This results into a significant improvement. For the above example involving a linear interaction, it is 9.8236, very close to the exact 9.8346. At the optimum, the effective masses are  about $\mu_{Mm}\sim 4$ and $\mu_{mm}\sim 2$. Once more, the optimized bound is seen to redistribute the inertia over the different pairs. 

In \cite{1978JMP....19.1969H}, Hall studied the case of a finite mass for the center, i.e., the configurations of the type $(M,m,m,m,\ldots)$, with coordinates $\vi r0, \vi r1,\ldots$. With the Jacobi variables $\vec R$ for the center of mass and $\vi xi=\vi ri-\vi r0$, and their conjugate momenta $\vec P$ and $\vi qi$ with $i\ge1$, the kinetic energy operator $K$ reads
\begin{equation}
K=\frac{\vs P}{2(M+ N\,m)}+\sum_{i=1}^N \frac{\vis qi}{2}\left(\frac1M+\frac1m\right)+\sum_{1\le i<j\le N} \frac{\vi qi.\vi qj}{M}~,
\end{equation}
as customary in atomic physics, when one exhibits the so-called mass-polarization terms. 
In the ground-state of the whole system,  all $\vis qi$ have the same expectation value, as well all $ \vi qi.\vi qj$. Hence the expectation value of the $(N+1)$-body Hamiltonian within its translation-invariant ground state can be seen as $N/2$ times the expectation value of the 3-body Hamiltonian
\begin{equation}
H_3'=\frac{\vis q1+\vis q2}{2}\left(\frac1M+\frac1m\right)+(N-1) \frac{\vi q1.\vi q2}{M}+V_{01}+V_{02}+(N-1)\,V_{12}~,
\end{equation}
which is, in the mass-polarization representation, the intrinsic part of an Helium-like Hamiltonian with potentials  $V_{01}$,  $V_{02}$, and $(N-1)\,V_{12}$, and masses $\{M',m',m'\}$, with $M'=M/(N-1)$ and $m'=(1/m-(N-2)/M)^{-1}$. 

For an overall harmonic interaction $V_{ij}(r)=r_{ij}^2$, and $N=3$, and masses $M=15$ and $m=1$, one gets a lower bound $E_4\ge 10.36$, while the exact value is $10.81$. In this case, the optimized bound of Sec.~\ref{subse:opt4} is exact. 

For a linear interaction $V_{ij}(r)=r_{ij}$, and same masses, the lower bound is $E_4\ge 9.42$, not very close to the exact $9.97$. This is better than the simple improved bound of Sec.~\ref{subse:imp-uneq}, which is  8.84, but significantly worse than the optimized bound $9.96$. 

\section{Application to Borromean binding}\label{se:coup:thr}
\subsection{Borromean binding of three or more bosons}
So far, we compared the energies of the $N$-body and $N'$-body systems at given coupling. Another point of view consists of comparing the couplings corresponding to a vanishing energy, i.e., the coupling thresholds.  It is well known that for a number of dimensions $d>2$, a minimal coupling is necessary for binding in a short-range attractive potential. Starting with the seminal paper by Thomas \cite{1935PhRv...47..903T}, it is also observed that the coupling threshold required to bind a three-body system, $g_3$, is  smaller than the one, $g_2$, of a two-body system.\footnote{The convention here for  $g\,V$ is that  $g>0$ and $V$ contains negative parts.}\@ This phenomenon is now referred to as Borromean binding. 

From \eqref{eq:naive}, one gets a crude limit $g_3\ge g_2/2$ on the window for Borromean binding. It can be refined to 
\begin{equation}\label{eq:Fleck-R}
 g_3\ge 2\,g_2/3~,
\end{equation}
from \eqref{eq:improved}, which is saturated for a harmonic well made vanishing at very large distances, for instance
\begin{equation}
 V(r)=(-A+r^2)\exp(-\lambda\,r)~,\quad A>0~,\quad \lambda>0~
\end{equation}
in the limit $\lambda\to 0$. See, for instance, \cite{1994PhRvL..73.1464R,2000PhRvA..62c2504M}. For simple monotonic potentials such as the ones with a Gaussian, or exponential or Yukawa shape, one gets typically $g_3/g_2\simeq 0.80$. 

The inequality \eqref{eq:Fleck-R} is easily generalized for $N$ bosons as 
\begin{equation}
 N\,g_N \ge (N-1)\,g_{N-1}~,
\end{equation}
i.e., $N\,g_N$ is an increasing function of $N$.  For the decreasing character of $g_N$, see, e.g., \cite{2003JPhA...36.6725G}.
\subsection{Borromean binding of three distinguishable particles}
For non-identical particles, the inequality involves several couplings. If, for instance, ones considers a system $(a,a,b)$ with two couplings normalized such that $g_{ij}$ is the coupling threshold for binding the $\{i,j\}$ pair, one can distinguish two domains in the $\{g_{aa}, g_{ab}\}$ plane, one near  $g_{aa}=g_{ab}=0$ without 3-body binding and another one with 3-body binding and four regions (see, e.g., \cite{2006FBS....38...57F}):
\begin{itemize}\itemsep -1pt
 \item for $g_{aa}>1$ and  $g_{ab}>1$, 3-body binding is obvious,
 \item for $g_{aa}<1$ and $g_{ab}>1$, one subsystem is unbound, this is ``tango binding'',
 \item for $g_{aa}>1$ and $g_{ab}<1$, two subsystems are unbound, this is ``samba binding'',
 \item for $g_{aa}<1$ and $g_{ab}<1$, the three  subsystems are unbound, this is ``Borromean binding'',
\end{itemize}

\begin{figure}[ht!]
 \centering
 \includegraphics[width=.35\textwidth]{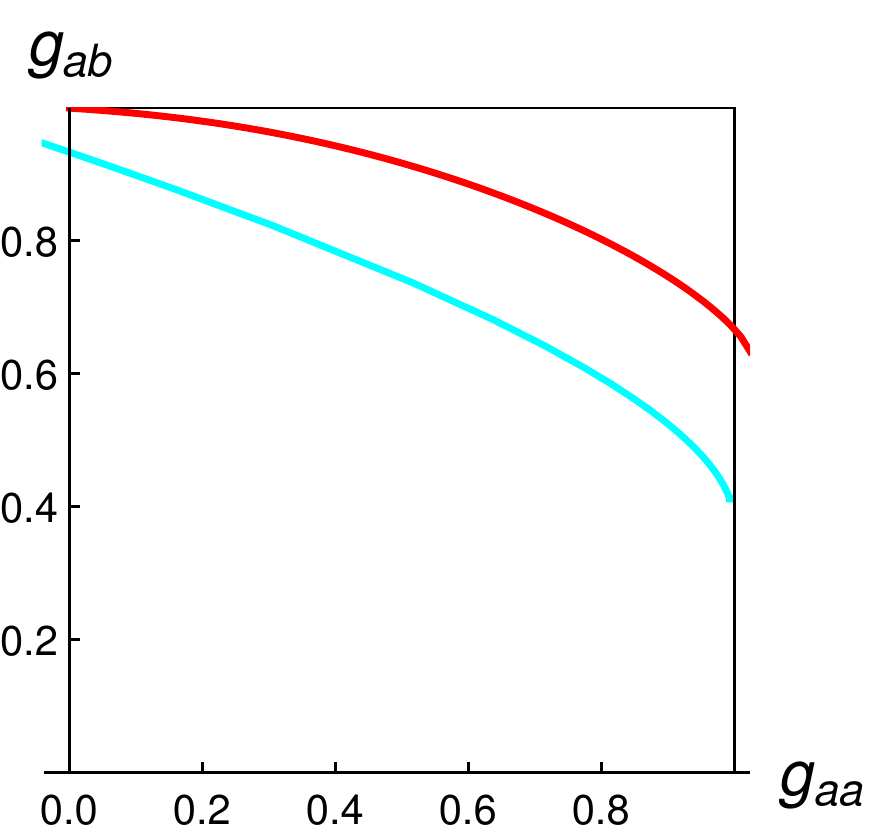}
 \caption{Strict limit for Borromean 3-body binding (cyan) of  the system $(a,a,b)$ with equal masses $m_a=m_b=1$, in the plane of the couplings $g_{aa}$ and $g_{ab}$ normalized such that $g_{ij}=1$ is the coupling threshold for the $\{i,j\}$ pair. The red curve is a numerical estimate in the case of an exponential interaction $-g_0^{-1}\,\sum_{i<j} g_{ij} \,\exp(-r_{ij})$, with $ g_0\simeq  1.4458$.}
 \label{fig:coupling123}
\end{figure}

An example is given in Fig.~\ref{fig:coupling123}, with an exponential potential. There is a substantial domain of possible Borromean binding. 
\subsection{Borromean binding of four particles}
For $N\ge4$, one can envisage several definitions of a Borromean system, for instance by requiring that all subsystems are unbound. According to the prevailing convention, a system is Borromean if there is no path to build it by adding the constituents one by one. For instance, the positronium hydride $(p,e^+,e^-,e^-)$ is remarkable since the second electron stabilizes the unstable $(p,e^-,e^+)$, but it is not Borromean if one considers the chain $p\to (p,e^-)\to (p,e^-,e^-) \to (p,e^-,e^-,e^+)$. On the other hand for $M\simeq 2\,m$, the  purely Coulombic system $(M^+,M^-,m^+,m^-)$  is genuinely Borromean, as it is stable, while none of its 3-body subsystems is stable \cite{2003PhRvA..67c4702R}.

The case of four particles interacting through a short-range potential was studied in \cite{1995PhRvA..52.3511G}. For $(m,m,M,M)$ with two pairs of bosons (or fermions in a spin singlet state), there are three couplings, $\{g_{mm},\, g_{MM}, \,g_{mM}\}$, normalized such that $g_{ij}=1$ is the critical coupling for binding in the attractive 
potential experienced by masses $m_i$ and $m_j$, so that the Hamiltonian reads
\begin{equation}
 H_4=\frac{\vis p 1+\vis p 2}{2\,m}+\frac{\vis p 3+\vis p 4}{2\,M}+\frac{g_{mm}}{m} u_{12}+\frac{g_{MM}}{M} v_{34}+\frac{(M+m)\,g_{mM}}{2\,m\,M}\sum_{i=1,2\atop j=3,4} w_{ij}~.
\end{equation}

\begin{figure}[h!!]
 \centering
 \includegraphics[width=.3\textwidth]{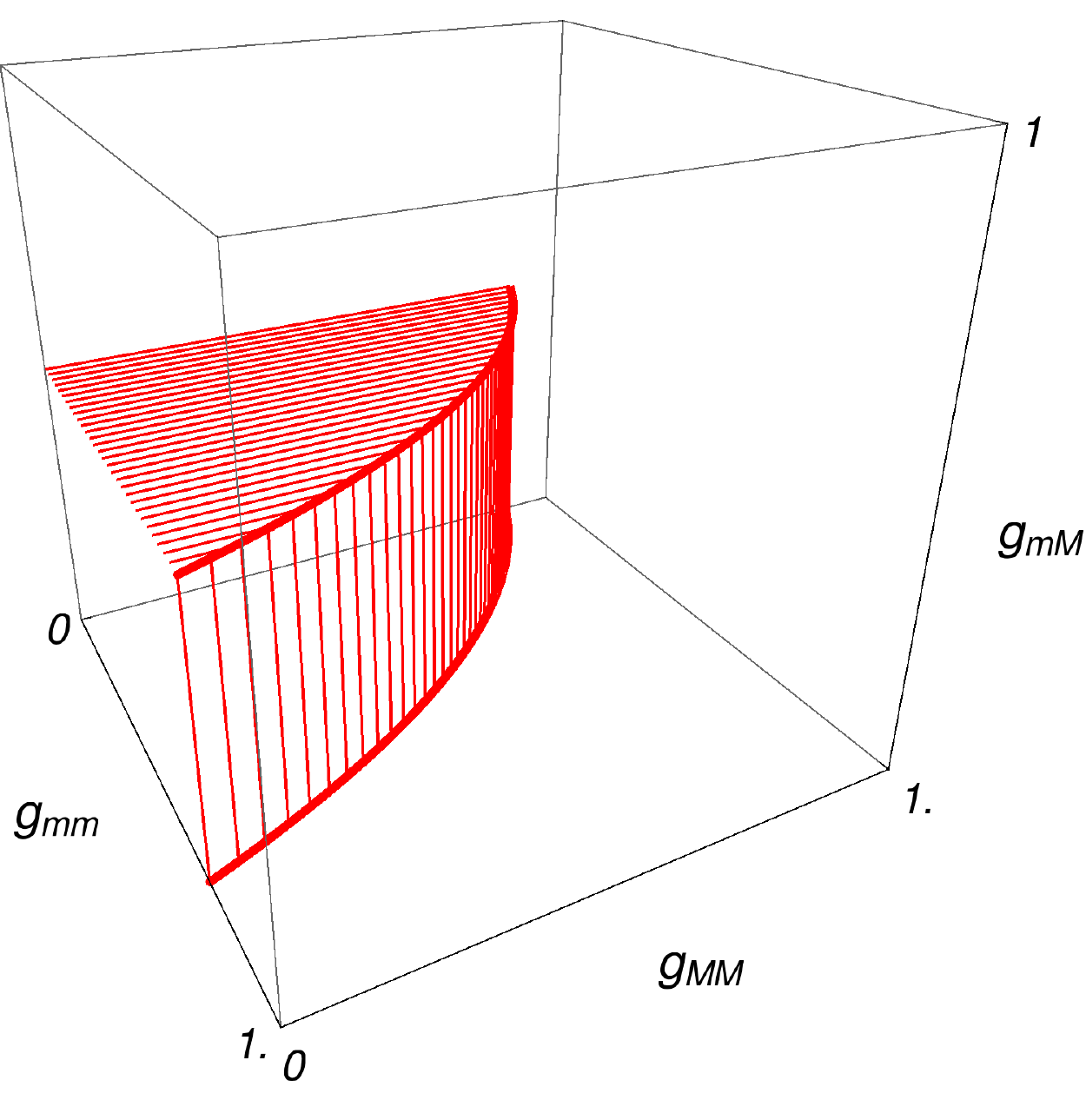}
 \quad
 \includegraphics[width=.3\textwidth]{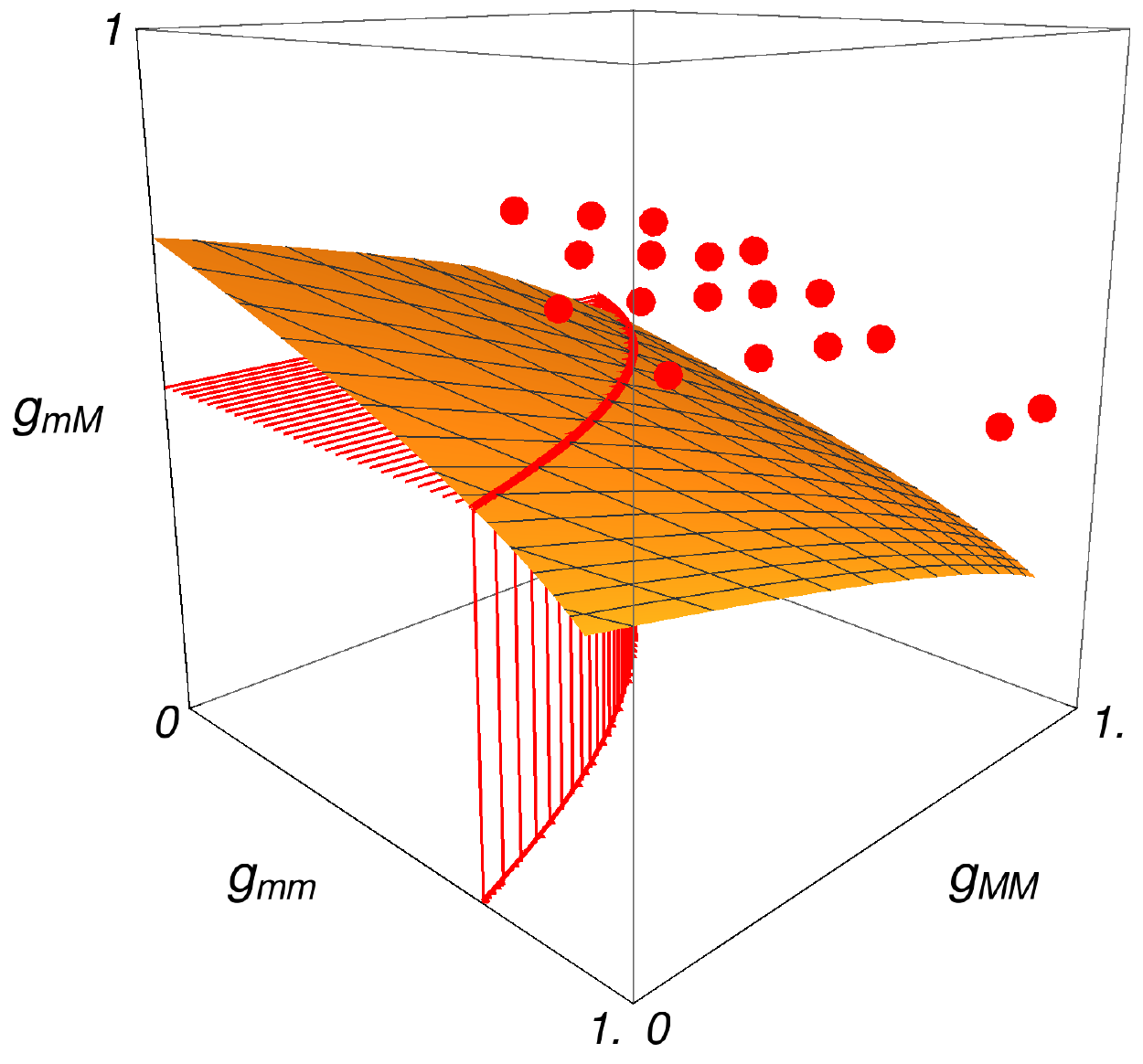}
 \quad
 \includegraphics[width=.3\textwidth]{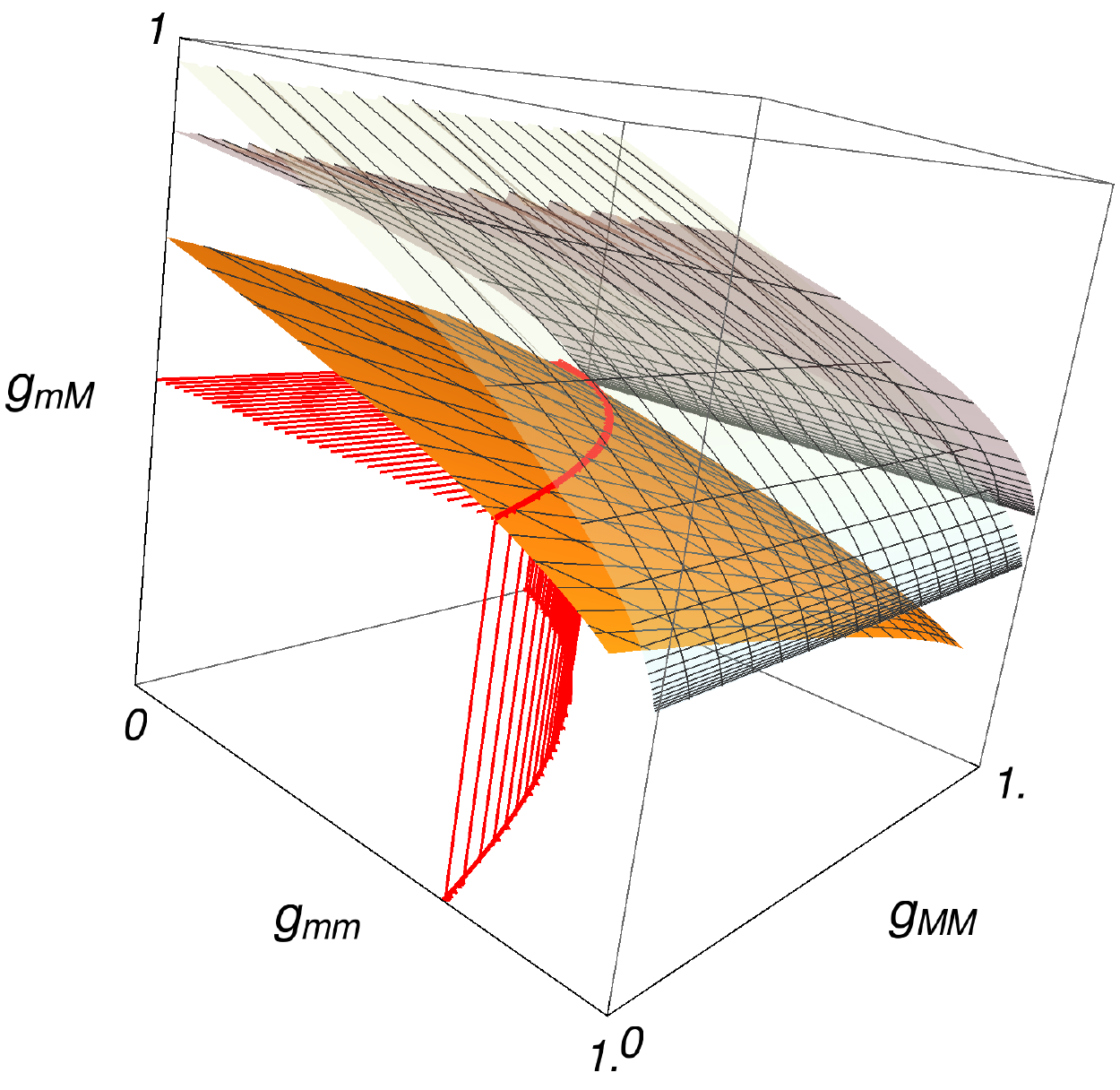}
 \caption{Forbidden domain of binding for $(m,m,M,M)$ with $M/m=2$, using the partially optimized decomposition \eqref{eq:4part-part-opt} (left), or the fully optimized decomposition \eqref{eq:opt-4masses} (center).  Also shown (right) are the forbidden domains for $(m,m,M)$ and $(M,M,m)$ binding. The red points in the figure in the middle correspond to an estimate of the critical coupling for a potential $\propto \exp(-r^2)$.}
 \label{fig:coupling4from2}
\end{figure}

In \cite{1995PhRvA..52.3511G}, the decomposition of the kinetic part as
\begin{multline}\label{eq:4part-part-opt}
 T_4=(\vi p1 + \vi p2+ \vi p3+\vi p4).(a\,\vi p1+ a\, \vi p2+ A\,\vi p3+ A\,\vi p4)\\
 {}+\frac x4(\vi p1-\vi p2)^2+\frac y4(\vi p3-\vi p4)^2+\frac z4\,\sum_{i=1,2\atop j=3,4} \left((1-\alpha)\vi pi-\alpha\,\vi pj\right)^2~,
\end{multline}
led to the conclusion that binding cannot happen if the inequalities
\begin{equation}
 \begin{aligned}
  g_{mm}&< - (m/M )^2\,\alpha^2+1-\alpha^2~,\\
g_{MM}&<- (M/m)\,(1-\alpha)^2 +\alpha\,(2-\alpha)~,\\
g_{mM}&<1/2~,
 \end{aligned}
\end{equation}
are simultaneously  satisfied. For $M/m=2$, after optimization of the parameter $\alpha$, this corresponds to the cylindrical section shown in Fig.~\ref{fig:coupling4from2} (left), while the fully optimized decomposition \eqref{eq:opt-4masses} extends significantly the forbidden domain, as seen in the plot at the center.  Also shown (right) is the volume where 4-body binding is possible while 3-body binding is strictly forbidden. Of course, the  domain of couplings for which 4-body binding and 3-body binding actually occur, depends on the details of the potentials. 


%
\section{Application to few-charge systems}\label{se:few-charges}
Systems made of three or four unit charges, such as $\mathrm{H}^-(p,e^-,e^-)$ or $\mathrm{Ps}_2(e^+,e^+,e^-,e^-)$, have been thoroughly studied. In this section, we briefly review the lower bounds of Hall-Post type. This illustrates the hierarchy among the various bounds, and the difficulty to get any accurate lower bound when one or several pairs do not support a bound state. In short, varying the attraction in a pair that binds will modify both the exact energy and the HP bounds.  Increasing the repulsion in a pair, on the other hand, will lower the binding but keep unchanged the HP bounds written in terms of 2-body energies. 
\subsection{Three unit charges}
We restrict here to configurations of the type $(M^\pm,M^\pm,m^\mp)$, which extend from $\mathrm{H}_2{}^+$ for $M\gg m$ to $\mathrm{H}^-$ for $M\ll m$, including the positronium ion for $M=m$.  For any $M$ and $m$, these systems are stable below the dissociation threshold $(M^\pm,m^\mp)+M^\pm$~\cite{1977JMP....18.2316H}, whose energy is $E_\text{thr}=-M\,m/(2(M+m))$.  The naive and improved bounds are given by \eqref{eq:naive:uneq} and~\eqref{eq:improved-uneq}, respectively, with the prescription  $E_2=0$ for pairs with equal charges.  The optimized bound corresponds to a decomposition
\begin{multline} \label{eq:3unit1}
H_3=(\vi p1+\vi p2+\vi p3)(A\,\vi p1+A\,\vi p2+a\,\vi p3)\\
{}+ x_{12}^{-1}\genfrac{(}{)}{}{}{\vi p2-\vi p1}{2}^2
+ x_{13}^{-1}\left[ \genfrac{(}{)}{}{}{\vi p1-u\,\vi p3}{1+u}^2+1\leftrightarrow2\right]~.
\end{multline}
After identification, one gets the effective masses $x_{12}$ and $x_{13}$ as a function of the parameter $u$. The best lower bound is not an isolated maximum, but is obtained for $x_{12}$ infinite, and $x_{13}$ minimal in the range of $u$ that keeps $x_{12}$ positive. 

\begin{figure}[ht!]
\centerline{%
 \includegraphics[width=.49\textwidth]{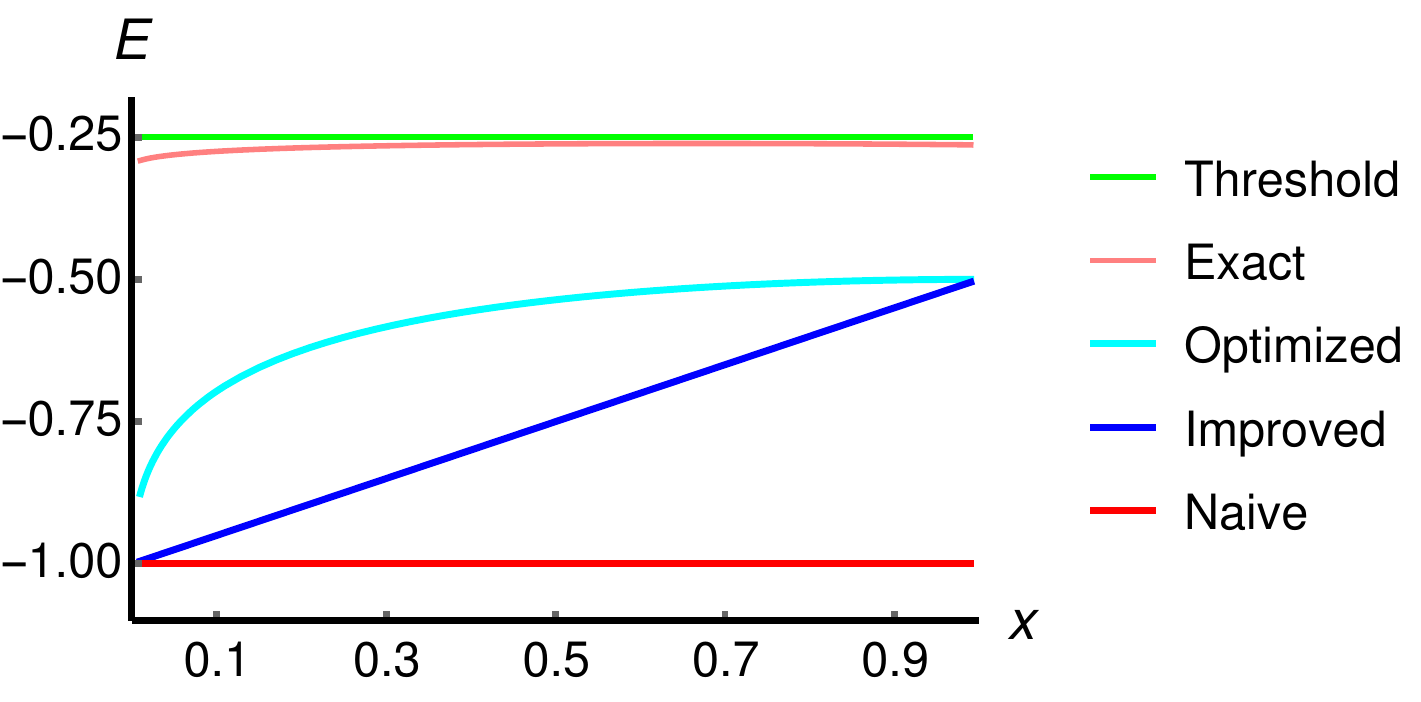}
\hfill
 \includegraphics[width=.49\textwidth]{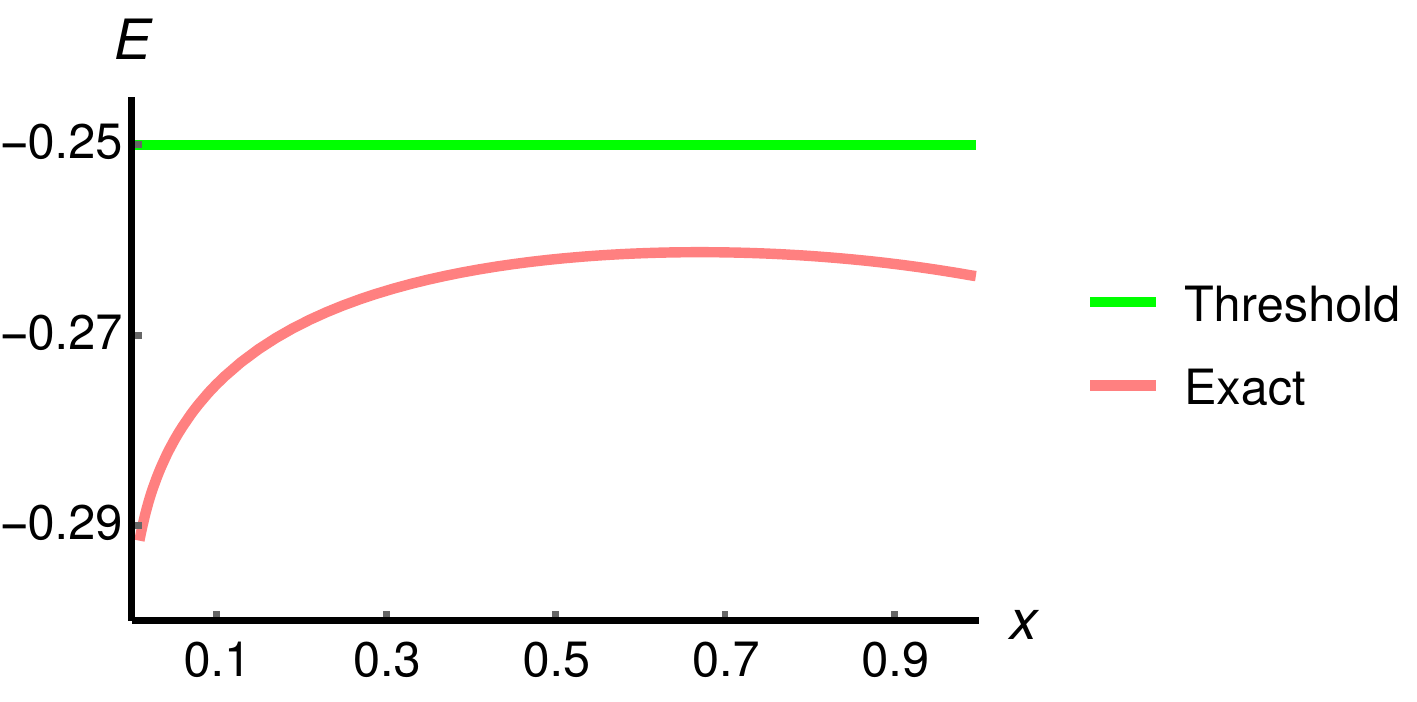}}
 \caption{Hall-Post bound for the energy of $(M^\pm,M^\pm, m^\mp)$ as a function of $x$,
 where $M^{-1}=2\,x$ and $m^{-1}=2\,(1-x)$, vs.\ the threshold energy and an accurate estimate of the ground-state energy.
 The  latter ones are distinguished in the second panel.}
 \label{fig:3charges}
\end{figure}

The various bounds are shown in the first panel of Fig.~\ref{fig:3charges} as a function of the ratio $x=m/(M+m)$ with $M^{-1}+m^{-1}=2$, so that the threshold energy is constant. In the second panel, an enhancement distinguishes the exact energy from the threshold.  The numerical estimate of the energy of $(M^\pm,M^\pm,m^\mp)$ as a function of the mass ratio $M/m$ will be used in the next subsection, as an input to the lower bound of $(M^+,M^+,m^-,m^-)$.  

These bounds on $(M^\pm,M^\pm,m^\mp)$ are rather poor, and even useless: for instance, a rigorous study of the stability of the positronium and hydrogen-like molecules requires beforehand to establish that the lowest threshold is made of two atoms, $(M^+,m^-)+(M^+,m^-)$, and not of an ion and an isolated charge, $(M^+,M^+,m^-)+m^-$ or $(M^+,m^-,m^-)\linebreak[0]{}+M^+$, i.e., that none of these  three-body ions are bound  by more than twice the $(M^+,m^-)$ energy.  This can be done \cite{1993PhRvL..71.1332R}, but not with the HP bounds shown in Fig.~\ref{fig:3charges}.
\subsection{Four unit charges}
We now consider systems of four unit charges, restricting to mass configurations $(M^+,\linebreak[0]M^+,m^-,m^-)$. As for $(M^+,M^+,m^-)$, the two-body type of decompositions are rather disappointing.  The naive bound is 
\begin{equation}\label{eq:naive-4}
 E_\text{nai}= -\frac{6\,m\,M}{m+M}~,
\end{equation}
which is 6 times the threshold energy
\begin{equation}\label{eq:thr-4}
 E_\text{thr}= -\frac{m\,M}{m+M}~.
\end{equation}
The improved bound is slightly better,
\begin{equation}\label{eq:imp-4}
 E_\text{imp}= -\frac{4\,m\,M}{m+M}~,
\end{equation}
but still much too far from an interesting approximation to the exact energy, which is  just below the threshold, even for $M\gg m$. 

\begin{figure}[ht!]
 \centerline{
 \includegraphics[width=.5\textwidth]{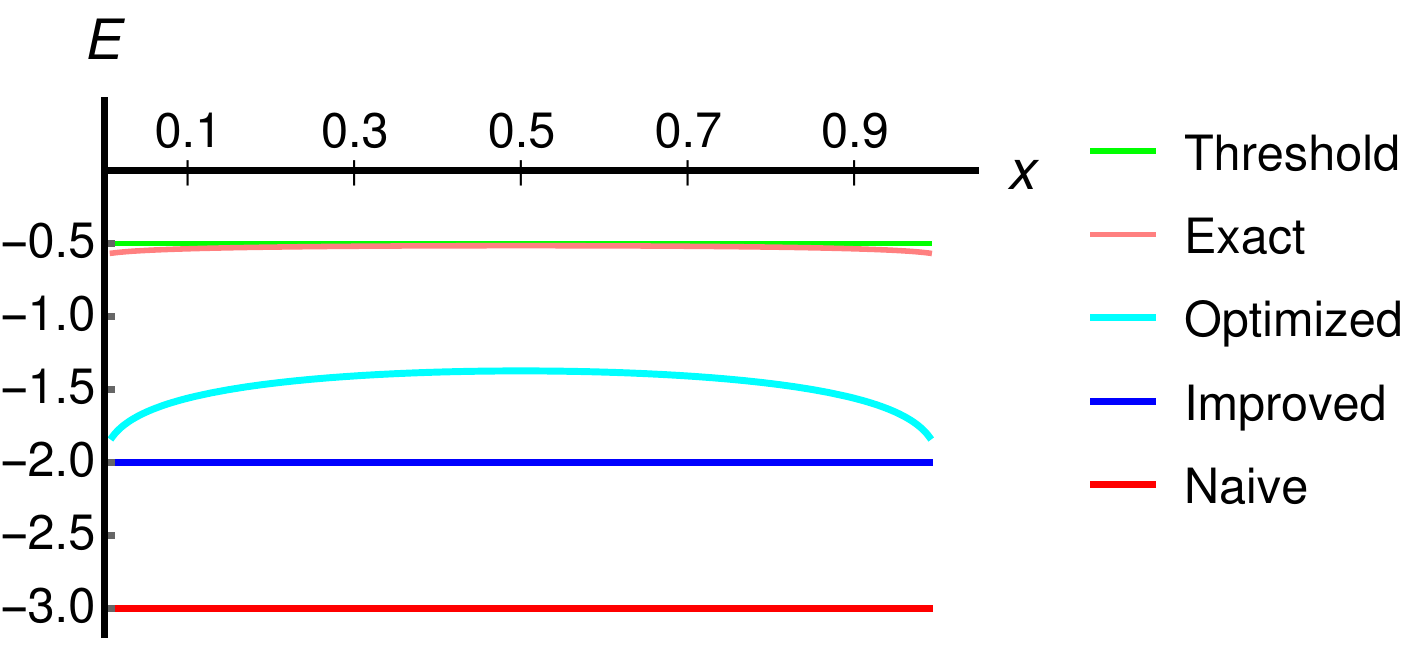}\hfill \includegraphics[width=.5\textwidth]{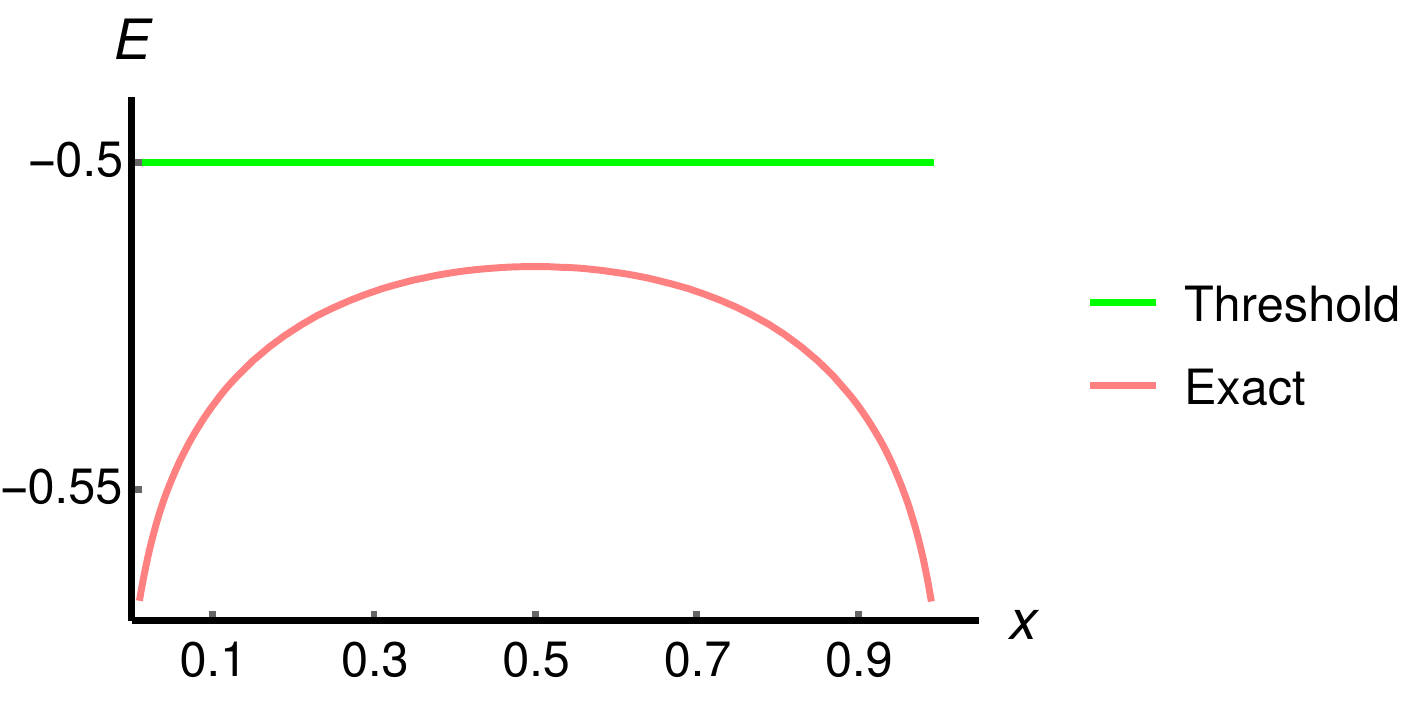}}
 \caption{Various HP lower bounds for the ground state of $(M^+,M^+,m^-,m^-)$, compared to the threshold and to the numerical estimate of the exact energy, which is magnified in the right panel.}
 \label{fig:MMmm-mol1}
\end{figure}

In the case of the $(M,M,m,m)$ mass distribution, the decomposition leading to the optimized bound reads 
%
\begin{multline}\label{eq:decomp-MMmm-coul}
H_4=(\vi p 1+\vi p2+\vi p3+\vi p4).(a\,\vi p1+a\,\vi p2+b\,\vi p3+b\,\vi p4)\\
{}+\frac{x}{4}\,(\vi p1 -\vi p2)^2+\frac{1}{r_{12}}+ \frac{y}{4}\,(\vi p3 -\vi p4)^2+\frac{1}{r_{34}}\\
{}+\sum_{\genfrac{}{}{0pt}{2}{i=1,2}{j=3,4}}\left[\frac{z}{4}\,(\vi p i -\vi p j+ c\, \vec p_{3-i} - c'\,\vec p_{7-j})^2-\frac{1}{r_{ij}}\right]~.
\end{multline}
After identification, one gets the three effective inverse masses $x$, $y$, $z$ (and  $a$ and $b$) as  functions of the parameters $c$ and $c'$, and the lower bound is given by
\begin{equation}\label{eq:imp-4a}
 E_\text{opt}=\max_{c,c'}\left[-\frac{1}{z(c,c')}\right]~,
\end{equation}
provided $x(c,c')>0$ and $y(c,c')>0$.  These latter constraints fix the maximum at
%
\begin{equation}\label{eq:imp-4b}
E_\text{opt}=8\left[ \sqrt{m+M} \left(\sqrt{m}+\sqrt{M}\right)- (m+M)-  \sqrt{m M}\right]~.
\end{equation}
 The various bounds are compared in Fig.~\ref{fig:MMmm-mol1}, where the threshold and the exact energy are also shown. The HP bounds are clearly very poor. 

It can be hoped that the repulsive character is somewhat ``digested'' in a decomposition  in terms of 3-body subsystems. 
For Ps$_2$, the naive bound is $E_\text{nai}=E(e^+,e^+,e^-)\simeq -0.786$, while the exact value is $E\simeq -0.516$. With the center-of-mass removed in the molecule and in the 3-body subsystems, one gets a slightly better  improved bound
$E_\text{imp}=8\,E(e^+,e^+,e^-)/3\simeq -0.699$. 

For the hydrogen like molecules $(M^+,M^+,m^-,m^-)$, one can use the formalism of Sec.~\ref{subse:4outof3}. The result is 
\begin{equation}
 E_\text{opt}=\max_{\tilde M} E_3(\tilde M^+, \tilde M^+,\tilde m^-)~,
\end{equation}
with 
\begin{equation}
 \tilde m^{-1}=\frac{1}{2} \left(1-2/\tilde M+\sqrt{1-2/\tilde M}\right)~.
\end{equation}
The maximum is reached for very large $\tilde M/\tilde m$, perhaps $\tilde M\to\infty$, but the precise determination of the maximum would require a detailed knowledge of the ground-state energy of $(\tilde M^+, \tilde M^+,\tilde m^-)$, for which only numerical estimates are available. We avoid here to start a meticulous interpolation of the energies computed by various authors, and restrict to an estimate based on the $M\to\infty$ limit of the hydrogen molecular ion
\begin{equation}
 E_\text{opt}\simeq  E(\mathrm{H}^{\infty}_ 2{}^{\ +})\simeq- 0.6~.
\end{equation}
It is intriguing that when optimizing the lower bound of the very  symmetric Ps$_2$, one reaches  the maximal asymmetry of the subsystems.

The exercise is now repeated for a hydrogen-like $(M^+,M^+,m^-,m^-)$ with $m=1$ and $M=3$, with threshold $E_\text{thr}=-3/4$ in our units. 
After a removal  of the center-of-mass energy in the molecule and subsystems, one gets
\begin{equation}
 E_\text{imp}=E(m_1^+m_1^+m_2^-)/2+E(m_3^+m_3^+m_4^-)/2=-1.085~,
\end{equation}
with $\{m_i\}=\{48/7,16/7,16/5,48/5\}$. If one now uses Eq.~\eqref{eq:E4from3-2} with $u=0$, then, from a rough estimate of the 3-body energy of the $(\tilde M^+, \tilde M^+,\tilde m^-)$ ions, one gets a maximum at $E_\text{opt}=-0.907$ with the masses $\{m_i\}=\{110., 1.49,6., 2.6\}$. If one freezes out these masses and introduces an asymmetry in the potential, through the parameter $u$ in \eqref{eq:E4from3-2}, the 4-body energy is bounded below by the energies of the systems $(m_1^{+1},m_1^{+1},m_2^{-1-u})$ and $(m_3^{-1},m_3^{-1},m_4^{1-u})$, whose maximum is about $-0.906$ for $u\sim 0.2$. Of course, a small gain could be expected by tuning the masses $m_i$ and the asymmetry parameter simultaneously. 
\section{Application to baryons in the quark model}\label{se:mes-bar}
In simple quark models, there is a color factor $\tilde\lambda_i.\tilde\lambda_j$ in front of the pair  potential, which means that the interaction is assumed to be mediated  by the exchange of a color octet. Here $\tilde\lambda$ denotes the set of eight SU(3) generators, the analogs of the Pauli matrices for SU(2). 
More precisely, when the spin-spin term is included, the interquark potential reads
\begin{equation}\label{eq:pot-col-spin}
 V=-\frac{3}{16}\,\sum_{i<j}\llcol{i}{j}\left[v_\text{c}(r_{ij})+ \ssspin{i}{j}\,v_\text{ss}(r_{ij})\right]~,
\end{equation}
where the normalization is such that the bracket is the quarkonium potential. 
The color factor is twice larger for a quark-antiquark pair forming a color singlet than for a quark-quark pair forming an antitriplet. 

The simplest inequality is obtained when the spin-spin interaction is omitted.  An obvious variant of \eqref{eq:naive} is
\begin{equation}\label{eq:MtoB1}
 E_3(m;V/2)\ge \frac32\,E_2(m;V)~
\end{equation}
which immediately relates meson and baryon ground-states with a single quark mass. After adding the constituent masses, one gets an interesting relation between the masses per quark  of  mesons and  baryons~\cite{1982PhRvD..25.2370A,2002PhR...362..193N},
\begin{equation}\label{eq:MtoB2}
 \frac{\mathcal{M}(qqq)}{3}\ge\frac{\mathcal{M}(q\bar q)}{2}~.
\end{equation}
For instance, baryon-antibaryon annihilation through quark rearrangement into three mesons is energetically possible.  The generalization to unequal quark masses is straightforward \cite{1984PhLB..139..408R},
\begin{equation}\label{eq:MtoB3}
2\, \mathcal{M}(q_1q_2q_3)\ge \mathcal{M}(q_1\bar q_2)+\mathcal{M}(q_2\bar q_3)+\mathcal{M}(q_3\bar q_1)~.
\end{equation}

When the spin-spin interaction is restored, the relation also holds between a spin 3/2 baryon and a spin triplet meson, if the spin-spin force contains a factor $\tilde\lambda_i.\tilde\lambda_j\, \vec\sigma_i.\vec\sigma_j$, as in the class of models \eqref{eq:pot-col-spin},  since $\vec\sigma_i.\vec\sigma_j=1$ for aligned spins.  In the experimental spectrum, the relation is satisfied by the $\Omega^-$ and the $\phi$ (1604\,MeV vs.\ 1020\,MeV). To the extent that the spin-spin force is treated at first order, one can compare the spin-averaged baryons and spin-averaged mesons. 

\begin{figure}[ht!]
 \centering
 \includegraphics[width=.5\textwidth]{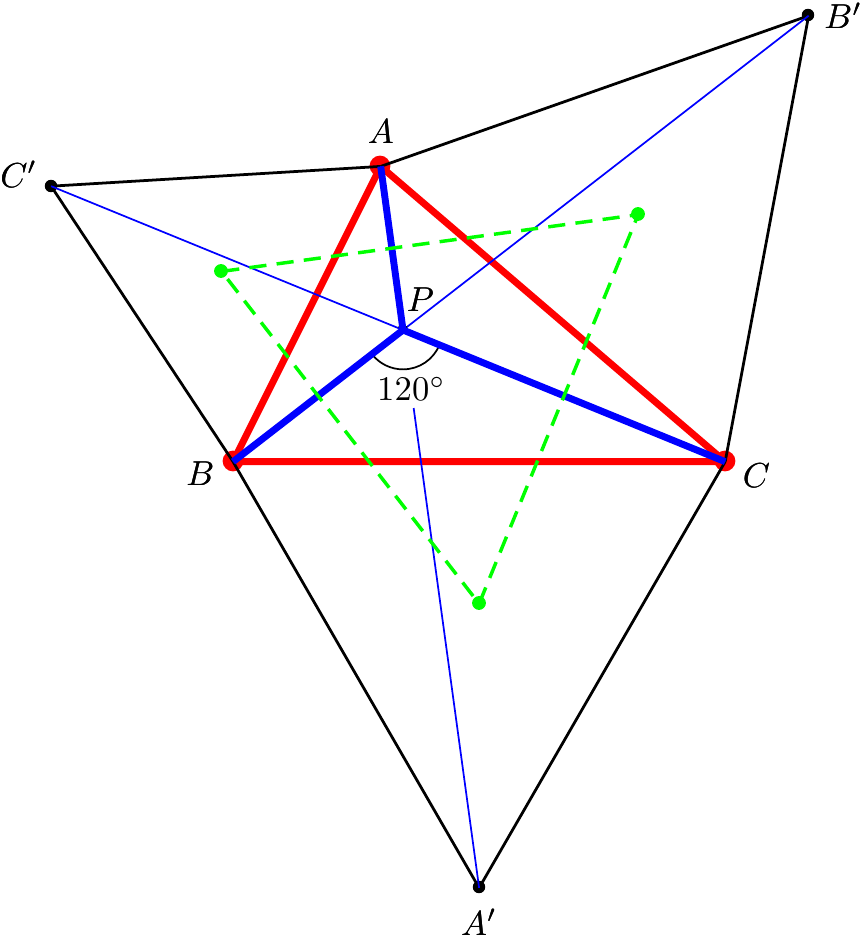}
 \caption{Construction of the minimal distance $L_\text{min}$ entering the confinement of three-quarks in a baryon. The minimal distance $L_\text{min}=PA+PB+PC$ is given by $L_\text{min}=AA'=BB'=CC'$, where $A'$ makes an equilateral triangle external to the quark triangle $ABC$, and so on. The Napol\'eon theorem states that the centers of the triangles $BCA'$, $CAB'$ and $ABC'$ form an equilateral triangle. }
 \label{fig:FermatTN}
\end{figure}

If one has in mind a potential of Cornell type, $V(r)=-a/r+b\,r$, with the 1/2 rule relating the meson and  baryon potentials, the above relations are automatically satisfied. If one now replaces for  the linear part the 1/2 rule by the prescription of the string model,
\begin{equation}\label{eq:MtoB4}
\begin{gathered}
 V_\text{meson}=-\frac{a}{r}+b\,r\\
 V_\text{baryon}=-\sum_{i<j}-\frac{a}{2\,r_{ij}}+b\,\min_J(r_{1J}+r_{2J}+r_{3J})~,
 \end{gathered}
\end{equation}
the inequality \eqref{eq:MtoB3} is reinforced as  the Fermat-Torricelli minimal distance obeys \cite{1976NuPhB.116..470D}
\begin{equation}
 L_\text{min}=\min_J(r_{1J}+r_{2J}+r_{3J})\ge (r_{12}+r_{23}+r_{31})/2~.
\end{equation}
The confining potential \eqref{eq:MtoB4} has been suggested more or less independently by several authors, so we refrain from listing an exhaustive set of references. The first paper is seemingly due to Artru \cite{1975NuPhB..85..442A}. The motivations leading to the minimal cumulated length \eqref{eq:MtoB4} are diverse: strong coupling regime, flux-tube picture, adiabatic limit of the bag model, etc. 
The first explicit quark-model calculations of baryon masses using this string potential suffered from the unjustified belief that one has first to determine the location of the junction $J$ and then compute $L_\text{min}=r_{1J}+r_{2J}+r_{3J}$. In fact, either by elementary geometry or by the general Melzak algorithm,\footnote{The special case when one of the angles of the $q_1q_2q_3$ triangle is larger than $120^\circ$ should be treated separately.} one can establish that $L_\text{min}$ is the distance from say, the quark $q_1$ to the auxiliary point $q'_1$ that makes an  equilateral triangle $q_2q_3q'_1$ external to the quark triangle $q_1q_2q_3$, as shown in Fig.~\ref{fig:FermatTN}. See, e.g., \cite{2009PhLB..674..227A}. Amazingly, the point $q'_1$ and its analogs $q'_2$ and $q'_3$ constitute the basis of a famous theorem by Napol\'eon \cite{coxeter1967geometry,laville1998geometrie}. 

With the advent of the improved bound \eqref{eq:improved}, the energy of a symmetric baryon with quark masses $m$ is bounded below by the energies of fictitious mesons with constituent masses $3\,m/4$. The increase of energy, from $E_2(m)$ to $E_2(3\,m/4)$ can be related to the observed excitation energy of quarkonia, with mild assumptions on the quark-antiquark potential.  See \cite{1990NuPhB.343...69B}.

A baryon of spin 3/2 can be compared to vector mesons via  \eqref{eq:MtoB3} for unequal masses, which reduces to  \eqref{eq:MtoB2} for equal masses. For a baryon with spin 1/2, one should make some averaging among spin-singlet and spin-triplet pairs. Consider for instance the ground state of $\Lambda(uds)$, in the isospin limit, i.e., with two constituent masses $m_q$ and $m_s$. Then one can write \eqref{eq:MtoB3} with a pure singlet for the pair $\{1,2\}$, and for each of the $\{1,3\}$ and $\{2,3\}$ pairs, a fictitious meson with $g=\langle \ssspin{i}{j}\rangle =0$, and use the convexity inequality
\begin{equation}
 \mathcal{M}(g=0)\ge \frac12\left[  \mathcal{M}(g=-3) +  \mathcal{M}(g=1) \right]~,
\end{equation}
leading to 
\begin{equation}
 \mathcal{M}(uds)\ge \frac12 \left[\mathcal{M}(q\bar q)_{S=0}+\mathcal{M}(q\bar s)_{S=0}+\mathcal{M}(q\bar s)_{S=1}\right]~.
\end{equation}
But this is not fully rigorous: if the spin-spin term of \eqref{eq:pot-col-spin} is treated non perturbatively  in the Schr\"odinger equation, the energy of $(uds)$ contains a small component with $ud$ in a state of spin $S=1$ and orbital momentum $\ell=1$, which lowers its energy. 
\section{Tetraquarks in simple  quark models}\label{se:tetra}
We now consider the Hamiltonian
\begin{equation}\label{eq:H-tetra}
 H=\sum_{i=1}^4\frac{\vis p i}{2\,m_i}-\frac{3}{16} \sum_{i<j} \llcol{i}{j}\,V(r_{ij})~,
\end{equation}
where $V(r)$ can be taken for example as a Cornell-type of  potential $V(r)=-a/r+b\,r+c$, or some ad-hoc power-law $V(r)=A\,r^\alpha$ or logarithmic form $V(r)=A\,\ln( r/r_0)$. 

There are two independent ways of building a color singlet out of two quarks and two antiquarks. If one chooses  the diquark-antidiquark basis, the states are
\begin{equation}
 \TT=(\bar 3,3)~,\quad \MM=(6,\bar 6)~,
\end{equation}
where the notation $\TT$ (true) and $\MM$ (mock) is inherited from color chemistry \cite{1978PhLB...76..634C}. One can also use the singlet-singlet or octet-octet states in the $(q_1\bar q_3)$-$(q_2\bar q_4)$ basis or in the other quark-antiquark pairing, or use the non-orthogonal basis made of the two possible singlet-singlet states. 
\subsection{Tetraquarks with frozen color wave function}
\boldmath\subsubsection{Color state $\TT=\bar33$}\unboldmath
In the approximation of a frozen $\TT$-color wave function with the two quarks in a color $\bar 3$ state, the Hamiltonian \eqref{eq:H-tetra} reduces to 
\begin{equation}\label{eq:H-tetra-33}
 H_{\TT}=\sum_{i=1}^4\frac{\vis p i}{2\,m_i}+\frac12\left[V(r_{12})+V(r_{34})\right]+\frac14\sum_{\genfrac{}{}{0pt}{2}{i=1,2}{j=3,4}}V(r_{ij})~,
\end{equation}
to which the  Hall-Post type of techniques are directly applicable. $H_\TT$ is purely confining, and, as such, does not exhibit any threshold. The question is  whether the coupled-channel Hamiltonian \eqref{eq:H-tetra}, when treated at the approximation  \eqref{eq:H-tetra-33} of a frozen $\TT$ color  wave function, can bind below the energy corresponding to two mesons. 

In the case of equal masses $m_i=m$, the decomposition
\begin{equation}
 H_{\TT}=\frac12(h_{12}+h_{34})+\frac14\sum_{\genfrac{}{}{0pt}{2}{i=1,2}{j=3,4}} h_{ij}~,\quad
 h_{ij}=\frac{\vis p i}{2\,m}+\frac{\vis p j}{2\,m}+V(r_{ij})~,
\end{equation}
where $h_{ij}$ is the quarkonium Hamiltonian,  provides the constraint
\begin{equation}\label{eq:mmmm-naive}
 E_{\TT} \ge 2\,E_2(m)~,
\end{equation}
which prohibits binding from $H_{\TT}$ alone. This can be easily improved by removing the center-of-mass motion of the subsystems. The identity
\begin{equation}
 \tilde H_{\TT}=\frac12\left[\tilde h_{12}(m)+\tilde h_{34}(m)\right]+\frac14\sum_{\genfrac{}{}{0pt}{2}{i=1,2}{j=3,4}} \tilde h_{ij}(m/2)~,\quad
 \tilde h_{ij}(m)=\frac{1}{m}\,\genfrac{(}{)}{}{0}{\vi p j -\vi p i}{2}^2+V(r_{ij})~,
\end{equation}
gives the lower bound
\begin{equation}\label{eq:mmmm-improved}
 E_{\TT}\ge  E_2(m)+E_2(m/2)~, 
\end{equation}
which is better than \eqref{eq:mmmm-naive}, due to the decreasing character of $E_2(m)$. But it is not perfect, as it is not saturated in the case of harmonic forces. For $m=1$ and $V(r)=r^2$, it gives $3\,(1+2^{1/2})\simeq7.243$ vs.\ the exact $E_{\TT}=3\,(3^{1/2}+2^{-1/2})\simeq7.317$. To get saturation, one should improve the decomposition as
%
%
\begin{multline}\label{eq:mmmm-optimized}
 \tilde H_{\TT}=\left[\left(\frac{x}{4}\,(\vi p 1 -\vi p 2)^2+\frac12\,V(r_{12})\right)+ (1,2)\leftrightarrow(3,4)\right]\\
{}+\left\{\left[\left(\frac{y}{4}\,(\vi p 1 -\vi p 3+ c\, \vec p_{2} - c\,\vec p_{4})^2+\frac14\,V(r_{13})\right)+1\leftrightarrow2\right]+ 3\leftrightarrow4\right\}~,
\end{multline}
and in the harmonic-oscillator case, saturation is reached for a somewhat exotic looking $c=5-2\,\sqrt{6}$. 

In the case of a linear potential $V(r)=r$, the threshold corresponds to $2\,E_2$, where $E_2\simeq2.3381$. The naive bound $2\,E_2$ indicates the absence of binding. The improved version pushes the lower bound to $(1+2^{1/3})\,E_2\simeq 2.260\,E_2$. The more flexible \eqref{eq:mmmm-optimized} gives $E_{\TT}>2.282\,E_2\simeq 5.336$, to be compared to  the exact $E_\TT\simeq 5.342$.

In the case of two different masses, say $(Q,Q,\bar q,\bar q)=(M,M,m,m)$, the naive decomposition gives 
\begin{equation}\label{eq:MMmm:naive}
E_\TT(M,M,m,m)>\frac12\,E_2(m)+\frac12 E_2(m)+E_2(\mu)~,\qquad \frac1{\mu}=\frac1M+\frac1m~,
\end{equation}
which is below the threshold $E_\text{th}=2\,E_2(\mu)$, an illustration of a theorem by Bertlmann and Martin, and Nussinov, stating that, for a given (flavor independent) potential, the 2-body ground-state energy is a concave function of the inverse reduced mass~\cite{1980NuPhB.168..111B,2002PhR...362..193N}. Thus the naive bound  does not prohibit  binding for any $M>m$, and stability is, indeed, reached for large enough $M/m$, the critical value of $M/m$ depending on the potential. 

The improved bound  prohibits  binding near $M=m$, till $M\simeq 9.55\,m$ in the case of HO  (binding occurs actually at $x\simeq 304\,m$). It reads
\begin{multline}\label{eq:MMmm-improved}
\tilde H_\TT=\frac{1}{4\,(M+m)}\left[(\vi p1 -\vi p2)^2+(\vi p3 -\vi p4)^2+\frac{V_{12}+V_{34}}{2}\right]\\
+\sum_{\genfrac{}{}{0pt}{2}{i=1,2}{j=3,4}}\left[\frac{2\,(M+m)}{m\,M}\,
\genfrac(){}{0}{m\,\vi pi-M\,\vi pj}{M+m}^2
+\frac{V_{ij}}{4}\right]~,
\end{multline}
with the somewhat surprising occurrence of the same effective mass for the $(M,M)$ and $(m,m)$ pairs, which  penalizes this bound for very asymmetric tetraquarks with large $M/m$. 

The optimized bound makes use of the decomposition 
\begin{multline}\label{eq:MMmm-optimized}
\frac{\vis p 1+\vis p 2}{2\,M}+\frac{\vis p 3+\vis p 4}{2\,m}+\left[\frac{V_{12}}{2}+\cdots\right]-(\vi p 1+\vi p2+\vi p3+\vi p4).(a\,\vi p1+a\,\vi p2+b\,\vi p3+b\,\vi p4)\\
{}=\frac{x}{4}\,(\vi p1 -\vi p2)^2+\frac{V_{12}}{2}+ \frac{y}{4}\,(\vi p3 -\vi p4)^2+\frac{V_{34}}{2}\\
{}+\sum_{\genfrac{}{}{0pt}{2}{i=1,2}{j=3,4}}\left[\frac{z}{4}\,(\vi p i -\vi p j+ c\, \vec p_{i'} - c'\,\vec p_{j'})^2+\frac{V_{ij}}{4}\right]~,
\end{multline}
with, again, $ i+i'=3$ and $j+j'=7$.  Saturation is obtained in the HO case. For instance, for $m=1$ and $M=5$, this corresponds to effective masses 
$x^{-1}=7.5$, $y^{-1}=1.5$, $z^{-1}=4.2$, instead of 6, 6 and 10/3 for the improved bound.  The decrease of the second effective mass explains why the optimized bound is significantly higher than the improved one. 

\begin{figure}[ht!]
 \centering
 \includegraphics[width=.5\textwidth]{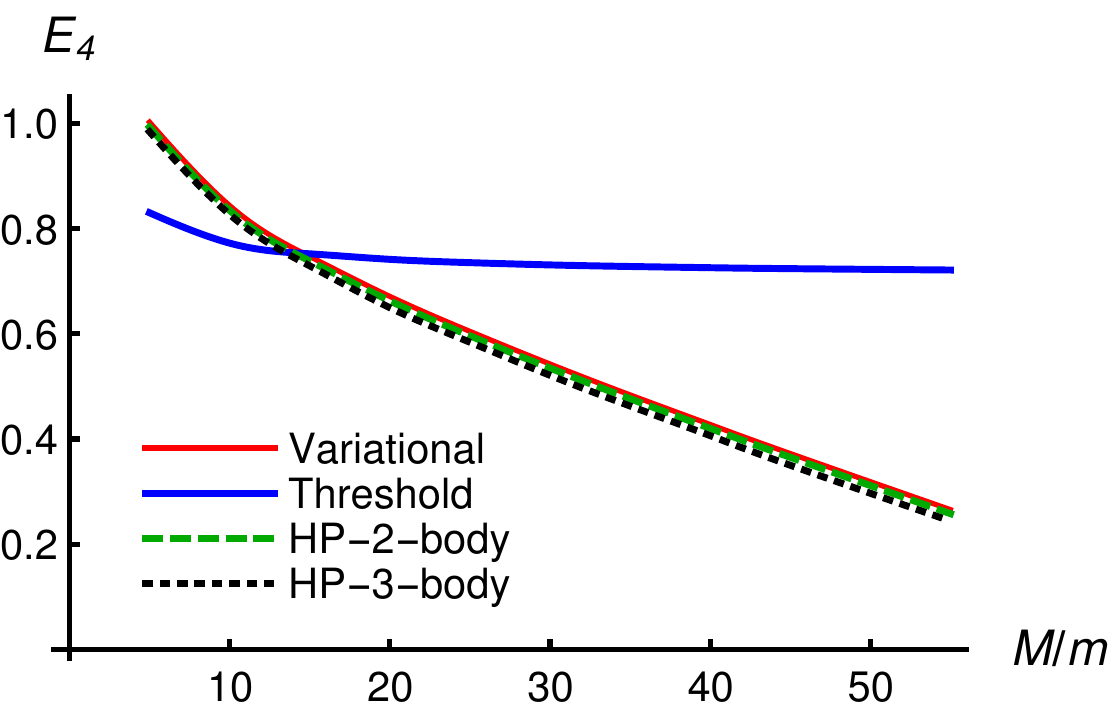}
 \caption{Variational energy vs.\ two variants of the Hall-Post lower bounds for the Hamiltonian $H_\TT$ of a tetraquark with frozen color $\TT=\bar33$, a decomposition into pairs, and a decomposition in 3-body clusters, namely \eqref{eq:MMmm-optimized}. Also shown is the threshold corresponding to two mesons. The light mass is taken as $m=1$. }
 \label{fig:TetraT}
\end{figure}

For illustration, we use a simple quark model, which reads
\begin{equation}\label{eq:Cornell-pot}
 V(r)=-\frac{0.4}{r}+ 0.2\,r~,
\end{equation}
with $V$ in GeV and $r$ in GeV$^{-1}$.  With a suitable additive constant, and heavy-quark masses $m_c=1.46$ and $m_b=4.85\,$GeV, it fits the first levels of $c\bar c$ and $b\bar b$.  In Fig.~\ref{fig:TetraT} are shown the  energy of the threshold $2\,(Q\bar q)$, the variational energy obtained with 8 generalized Gaussians,\footnote{A generalized Gaussian is an exponential in the most general quadratic polynomial of the Jacobi coordinates, supplemented by terms deduced by symmetry.} and the optimized Hall-Post bounds obtained from a decomposition either in 2-body or 3-body systems with $u=0$ in Eq.~\eqref{eq:E4from3-3}. The three energies are very close, and thus the exact result is very much under control.  

Note that the HP bound in terms of 3-body clusters is slightly below the one in terms of 2-body clusters,  contrary to what happens for equal masses with a symmetric interaction. If one restores the possibility of $u\neq 0$, then the 3-body HP bound is improved. In Table~\ref{tab:comp-mij}, we concentrate on the case $M/m=5$, with various degrees of sophistication for the HP bounds. The 3-body decomposition suffers from the fact that the quark-antiquark pairs appear in different clusters, $(QQ\bar q)$, and $(\bar q\bar q Q)$, with somewhat different effective masses for the $(Q\bar q)$ pairs. A comparison is done of these effective masses. 

\begin{table}[htb!]
 \caption{Comparison of the HP bounds for the $\TT$-tetraquark with masses $m=1$ and $M=5$ in the potential  \eqref{eq:Cornell-pot}.  The $m_{ij}$ are the effective masses of the pairs. For the decomposition HP-2, they are obtained directly. For the decomposition in terms of 3-body clusters,  the 3-body subsystems are, in turn, decomposed in 2-body clusters, and the $m_{ij}$ that are shown correspond to an average of the inverses, as each pair enters two 3-body clusters. The exact energy is $1.0020$. }
 \label{tab:comp-mij}
 \begin{center}
 \renewcommand{\arraystretch}{1.3}
  \begin{tabular}{ccccc}
   \hline\hline
   HP type & $E_\text{HP}$ & $m_{12}$ & $m_{34}$ & $m_{13}$\\
   \hline
   HP-2-imp &0.8721  &   6.00 & 6.00  &  3.33 \\
   HP-2-opt  & 0.9926  &   5.97  &   1.62   & 4.45 \\
   \hline
   HP-3-imp & 0.9744  &   5.90 & 2.19 & 4.05 \\
   HP-3-opt ($u=0$)&0.9837        & 6.74  &  1.69 &  4.24 \\
   HP-3-opt ($u\neq 0$)&0.9881    &  6.61 & 1.77 & 4.16 \\
   \hline\hline
  \end{tabular}
 \end{center}
\end{table}

\boldmath\subsubsection{Color state $\MM=6\bar6$}\unboldmath\label{sub:M-color}
For a $\MM=6\bar 6$ color state, the Hamiltonian is
\begin{equation}
 H_\MM=\sum_{i=1}^4\frac{\vis p i}{2\,m_i}-\frac14\left[V(r_{12})+V(r_{34})\right]+\frac58\sum_{\genfrac{}{}{0pt}{2}{i=1,2}{j=3,4}}V(r_{ij})~
\end{equation}

As explained, e.g., in \cite{2018PhRvC..97c5211R}, for equal masses $m_i=m$, the above $H_\MM$ is more favorable than $H_\TT$, due to the larger spread of the color coefficients around the same average value. For double flavor configurations $(M,M,m,m)$ with large $M/m$, the color state $\MM$, unlike $\TT$, does not benefit from the heavy-heavy attraction and gives a higher energy. 

As $H_\MM$ contains negative color coefficients, any decomposition in terms of 2-body sub-Hamiltonians would lead to a trivial $E_\MM>-\infty$. So one has to rely on a decomposition with 3-body subsystems. 
The simple identity
\begin{equation}
  H_\MM=\left[\frac{\vis p 1}{6\,m_1}+\frac{\vis p 2}{6\,m_2}+\frac{\vis p 3}{6\,m_3}-\frac18\,V_{12}+\frac5{16}\,V_{13}+\frac5{16}\,V_{23}\right]+\cdots
\end{equation}
with a summation over the missing quark or antiquark,  leads to a ``naive'' bound
\begin{equation}
  E_\MM>E_3(3\,m_1,3\,m_2,3\,m_3;-1/8,5/16,5/16)+\cdots~,
\end{equation}
where $E_3(m_1, \ldots;g_{12},\ldots)$ denotes the ground state of $H_3=\sum \vis p i /(2\,m_i)+\sum g_{ij}\,V_{ij}$. 
In the case of harmonic confinement, this naive bound allows binding for any $M/m$, even for $M=m$, in which case,  in units where $m=1$, $E_\text{nai}\simeq 5.968$ vs.\ $E_\text{th}=6$.  The improved bound, that uses the effective masses \eqref{eq:masses-imp}, prohibits binding near $M=m$. 
If one uses the Cornell type of potential \eqref{eq:Cornell-pot}, one gets, in the case of four equal masses $M$: for $M=m_c=1.46$, a lower bound to the tetraquark mass $M_\text{HP}=6 .973$ very close to the exact $M_\MM=6.978$; for  $M=m_b=4.85$, they become  $M_{HP}=19.725$ and  $M_\MM=19.735$.
\subsection{Tetraquarks with color mixing}
In the quark model \eqref{eq:H-tetra}, the wave function has two components, say
\begin{equation}\label{eq:psi-tetra}
 \Psi=\psi_\TT\,|\TT\rangle+ \psi_\MM\,|\MM\rangle~,
\end{equation}
and in this basis the potential reads
\begin{equation}
 \begin{pmatrix}
    \dfrac{V_{12}+V_{34}}{2}+\dfrac{V_{13}+V_{14}+V_{23}+V_{24}}{4} & \dfrac{3}{4\,\sqrt2}(V_{13}-V_{14}-V_{23}+V_{24}) \\[12pt]
    \dfrac{3}{4\,\sqrt2}(V_{13}-V_{14}-V_{23}+V_{24}) & -\dfrac{V_{12}+V_{34}}{4}+\dfrac{5\,(V_{13}+V_{14}+V_{23}+V_{24})}{8}	
   \end{pmatrix}~.
\end{equation}
The Hamiltonian now contains a continuum of states made of two well-separated color-singlets, starting from the threshold energy $E_\text{th}=2\,E_2(M,m)$, hence the HP lower bounds should satisfy $E_\text{HP}\le E_\text{th}$. In other words,  there is no hope to derive a lower bound that would exclude any binding. We shall get, however,  a lower bound close below the threshold, that  excludes the possibility of any deeply-bound state. 

We restrict the illustration to the case of equal masses, which corresponds to the speculations about $bb\bar b\bar b$, or its analog with charm quarks. 
This configuration is interesting, as it could be reached with a $\Upsilon\Upsilon$  or $\Upsilon\eta_b$ trigger, one of the quarkonia being virtual if the state is bound. Contrary to some naive belief, the heavy mass of the $b$ does not guarantee stability.  Increasing the constituent mass decreases the algebraic energy of both the tetraquarks and its dissociation products, and the net result is not obvious.  In current potential models, $bb\bar b \bar b$ is unstable \cite{Richard:2017vry,Chen:2019dvd,Liu:2019zuc}, contrary to some claims which suffer from an incorrect treatment of the 4-body problem, such as the use of a diquark-antidiquark approximation. 

There are two ground-state solutions of \eqref{eq:psi-tetra}. The first one is symmetric under the exchange of quarks or antiquarks in the $\MM$ sector, the second one in the $\TT$ sector.  For the  former  case, the variational solution of the coupled equations leads to a mass  $M_\text{var}=19.513$, slightly above the threshold $M_\text{thr}=19.501$, because we have not pushed the calculation far enough.  The state is clearly unbound, and indeed, the percentages of $\TT$ and $\MM$ components correspond to an almost exact singlet-singlet structure of color.  More interesting, after some optimization, the lower bound is obtained at about $M_\text{HP}=19.459$. This means that 3-body calculations exclude any 4-body binding exceeding $40\,$MeV.  For the other state, symmetric in the $\TT$ sector, one gets, of course the same threshold $M_\text{thr}$,  the variational $M_\text{var}=19.514$ and the lower bound $M_\text{HP}=19.480$. The results in the case where $M=m_c$ are very similar.  To our knowledge, this is the first use of HP inequalities with coupled channels.

\section{Miscellaneous}\label{se:misc}
Obviously, the survey of Hall-Post type of lower bounds could be developed in many aspects. Let us sketch two examples.
\subsection{Excited states}
In this review and in the literature, the Hall-Post inequalities were mainly discussed for the ground state, or in the case of fermions, the lowest state compatible with the requirements of statistics.  But there are a few exceptions.  

In \cite{1969PhLB...30..320H}, Hall noticed that \eqref{eq:N-fermions} provides a lower bound to excited states. If $\epsilon_0$, $\epsilon_1$, \dots, are the energy of the 2-body Hamiltonian with effective mass $2(N-1)m$ and coupling $g/2$, then the ground state of the $N$-boson system is bounded below by $N(N-1)\epsilon_0$ (as already noticed, this coincides with the naive HP bound), the first excited state by $N\min[(N-2)\epsilon_0+\epsilon_1]$, the second one by
$N\,\min[(N-2)\epsilon_0+\epsilon_2,(N-3)\epsilon_0+2\epsilon_1]$, etc. 

In a more recent paper, Hill \cite{1980JMP....21.1070H} devised a set of $n$ coupled integral equations, whose solution, as $n$ increases, improves the lower bound of the ground state, and provides also more and more stringent lower bounds for the excited states.  The example of the 3-body harmonic oscillator in 1 or 3 dimensions is treated in this paper, and a hard-core model was considered in \cite{1980JMP....21.1086B}.
Clearly more study is needed for excited states.
\subsection{Bounds for semi-relativistic Hamiltonians}
The naive decomposition \eqref{eq:naive-decomp} is independent of the form of the  kinetic energy operator $\vis pi/(2\,m_i)$ associated with each particle. We already mentioned the remark by Calogero and Marchioro \cite{1969JMP....10..562C}, that it remains valid if each one-body term is supplemented by an interaction with a fixed center, and becomes $\vis pi/(2\,m_i)+ g'\,U(r_i)$. 

Hall and Lucha \cite{2008JPhA...41I5202H,2009JPhA...42M5303H} studied the case where the non-relativistic energy operator is replaced by 
$(m_i^2+\vis pi)^{1/2}$.  It is straightforward to obtain (let us restrict for simplicity to the case of identical bosons, though some results are easily generalizable to unequal masses)
\begin{equation}
 H_3=\frac{1}{2}\,\sum_{i<j}\left[ \sqrt{m^2+\vis pi}+\sqrt{m^2+\vis pj}+2\,V(r_{ij})\right]~.
\end{equation}
It is notorious that  such relativistic form of energy does allow for an explicit separation of the center-of-mass kinetic energy, unlike the K\"onig theorem in the non-relativistic case. Nevertheless, as pointed out by Hall and Lucha, the relativistic energy is minimal in the rest frame where $\vi pi=-\vi pj$, so that there is a lower bound expressed in term of the eigenvalues of the single-variable operator  $(m^2+\vs p)^{1/2}+ g\, V(r)$. 
\section{Outlook}\label{se:conc}
We have presented a survey of the lower bound to the ground state energy of $N$-body systems given as a sum of energies of subclusters. In many cases, they offer a good approximation to the exact value. The techniques were developed first for three identical bosons, and extended to more particles and to unequal masses.  In the latter case, an essential progress was the so-called optimized bound. It is based on a decomposition of the Hamiltonian where, instead of a mere removing of the kinetic energy of the center-of-mass, one sets apart a more general term $(\vi p1+\vi p2+\cdots).(a_1\,\vi p1+ a_2\,\vi p2+\cdots)$ which also vanishes in the center-of-mass of the $N$-body system of interest. This provides more flexibility, and leads to inequalities that are saturated in the case of harmonic potentials. We have shown that the method of optimized bound gives better results than the ones in the literature. 

The decomposition of a $N$-body system in terms of $N'<N$ subsystems with $N'\neq 2$ was just a curiosity in the early literature. We have given examples of useful inequalities for $N'=3$, in cases where $N'=2$ is either useless or ill defined. This concerns hydrogen-like molecules in atomic physics or, in some quark models,  tetraquarks with two quarks in a color-sextet state.  The full treatment of tetraquarks  in such models requires an extension of the HP formalism  to include coupled channels. It provides useful restrictions on the possibility of binding full-heavy tetraquarks. 

The case of fermions remains  much more difficult than the one of bosons. Some progress has been achieved for an interaction close to the harmonic oscillator, thanks to subtle convexity inequalities. Another direction consists of analyzing the structure of the $N$-body wave function in terms of representations of the permutation group for its subsystems.  In the same vein is the extension of HP to radial excitations: the art of estimating a lower bound becomes more intricate than summing up sub-energies. The formalism will hopefully become more tractable in the future. 

To end up, we like to stress that deriving a good lower bound gives interesting insight on the $N$-body structures: how the interaction is shared among the clusters? which effective mass is acquired by each constituent inside each subcluster?  The same questions are of course raised when one tries to design an efficient and physically meaningful trial function for variational calculations. What can perhaps be looked for in the future  is a parallel and convergent development of a variational method in terms of clusters and a Hall-Post type of decomposition. 

%
%
%
%
%
%
%
\appendix
\section{Exact solution of the asymmetric oscillator}
We briefly recall the solution of the asymmetric oscillator, with different masses and strengths. Sometimes, one finds papers \cite{1996PhR...268..263G,2012EPJA...48...61Y}, where unnecessary approximations are done. Let 
\begin{equation}
 H=\sum_i \frac{\vis p i}{2\,m_i}+\sum_{\{i,j,k\}} g_k\,r_{ij}^2~,
\end{equation}
with, again, $\{i,j,k\}$ being an even permutation, and $r_{ij}=|\vec r_j-\vec r_j|$. A first change of variables, $\vec x_i=\sqrt{2\,m_i}\,\vec r_i$, leads to an operator 
\begin{equation}
 H=-\sum_i \frac{\partial^2}{\partial \vec x_i^2}+P~
\end{equation}
with a  symmetric form for the kinetic energy and a potential  $P$ that is a second-order polynomial in the $\vec x_i$. One then introduces three Jacobi coordinates with coefficients on the $x_i$. The third Jacobi variable corresponds to the center-of-mass $\vec R\propto \sum \vec x_i \,m_i^{1/2}$. For the two first ones, one simply imposes that the transformation $\{\vec x_i\}\to \{\vec u,\vec v,\vec R\}$ is orthogonal. The potential is independent of $\vec R$ and is a definite-positive function of $\vec u$ and $\vec v$. Thanks to a further orthogonal transformation, it is rewritten $P=\alpha\, \vec{u'}{}^2+\beta\, \vec{v'}{}^2$. Then the energy is $3(\alpha^{1/2}+\beta^{1/2})$.

A less pedestrian method consists of writing the potential as $\tilde r.A.r$, where $\tilde r=\{\vec r_i\}$ and $A_{11}=g_{2}+g_{3}$, $A_{12}=-g_3$, etc., and describing the transformation toward a symmetric form of kinetic energy by the diagonal matrix $B=\diag\{(2\,m_i)^{-1/2}\}$. Then the energy is simply
\begin{equation}
3\, \Tr(\tilde B.A.B)^{1/2}~,
\end{equation}
and there is no need for removing explicitly the center of mass, as it corresponds to  a vanishing eigenvalue. Most modern mathematical softwares include the computation of the square root and the trace of a matrix.  
\subsection*{Acknowledgements}
Past discussions of one of us (JMR) with J.-L.~Basdevant, A.~Krikeb, A.~Martin, Tai-T. Wu and S. Zouzou are gratefully acknowledged, as well as a correspondence with R.J. Carr. 
This work has been partially funded by Ministerio de Econom\'\i a, Industria y Competitividad (Spain)
and EU FEDER under Contract No. FPA2016-77177.
%

%
\end{document}